\documentclass[%
superscriptaddress,
nofootinbib,
 amsmath,amssymb,
 aps,
 longbibliography,
 twocolumn,
]{revtex4-2}

\usepackage{graphicx}
\usepackage{bm}
\usepackage{gensymb}
\usepackage{ragged2e}
\usepackage{mathtools}
\usepackage{lineno}
\usepackage[colorlinks=true, linkcolor = blue, urlcolor=blue, citecolor=red, anchorcolor = green]{hyperref}
\usepackage{newtxtext,newtxmath}
\usepackage{algorithm}
\usepackage[noend]{algorithmic}
\usepackage{subfigure}
\usepackage{adjustbox}
\usepackage{array}
\newcommand\vnimagef{\adjustbox{valign=m, vspace=0.1pt}{\includegraphics[width=.30\linewidth]{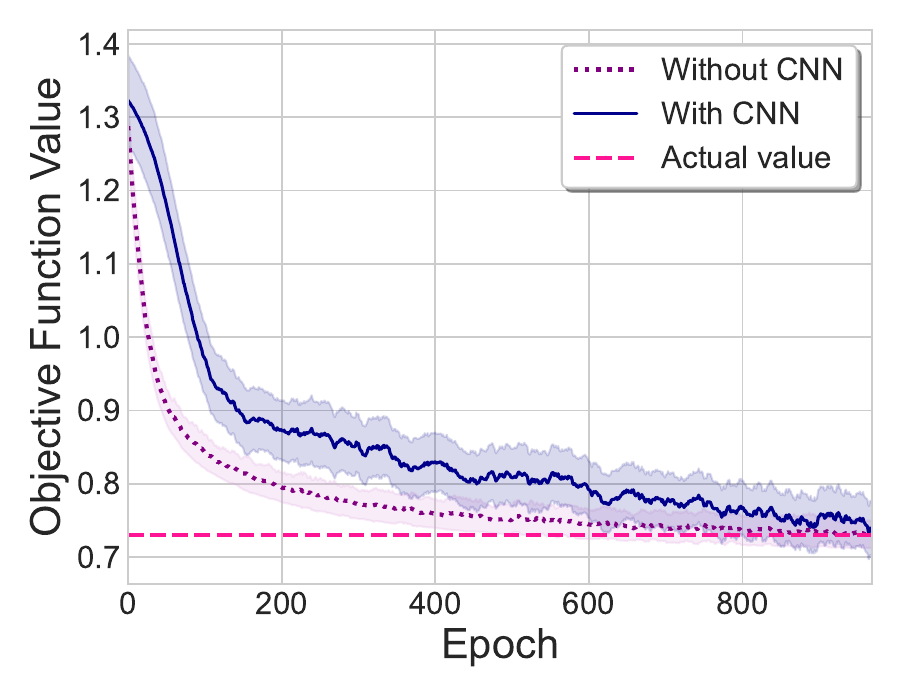}}}
\newcommand\vnimages{\adjustbox{valign=m,vspace=0.1pt}{\includegraphics[width=.30\linewidth]{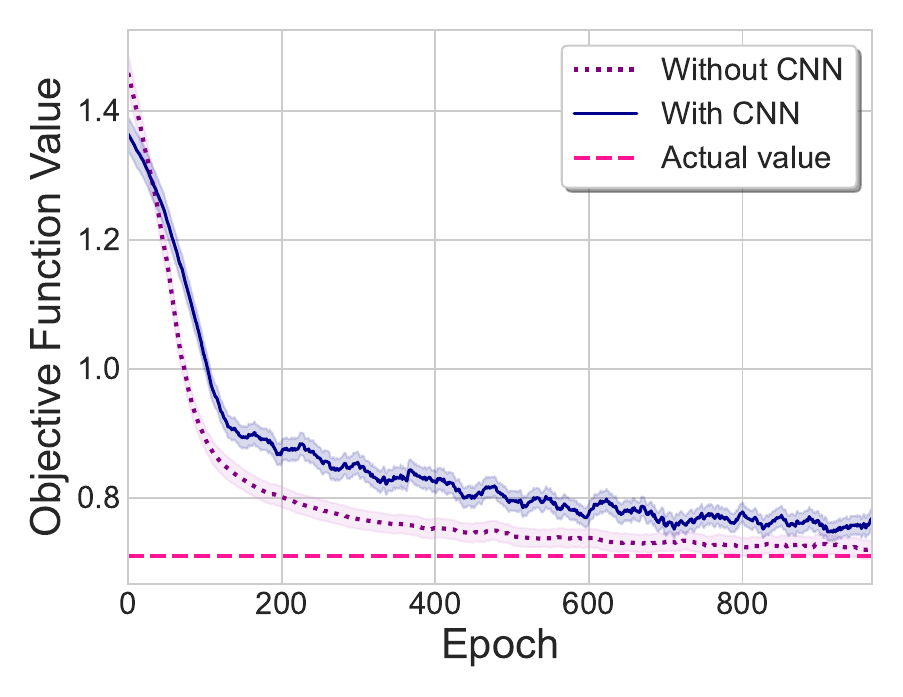}}}

\newcommand\fiimagef{\adjustbox{valign=m,vspace=0.1pt}{\includegraphics[width=.30\linewidth]{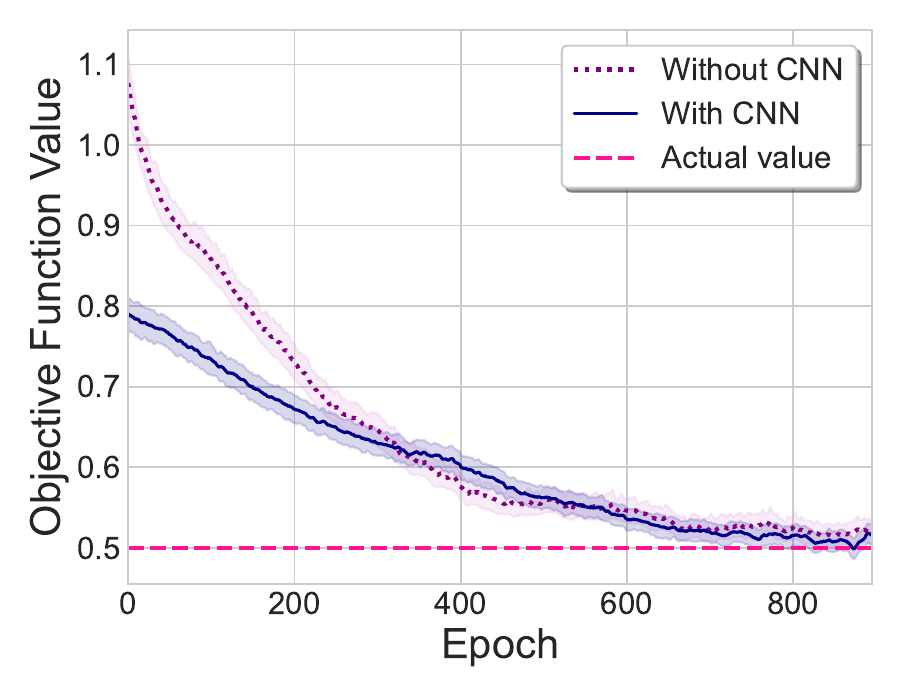}}}
\newcommand\fiimages{\adjustbox{valign=m,vspace=0.1pt}{\includegraphics[width=.30\linewidth]{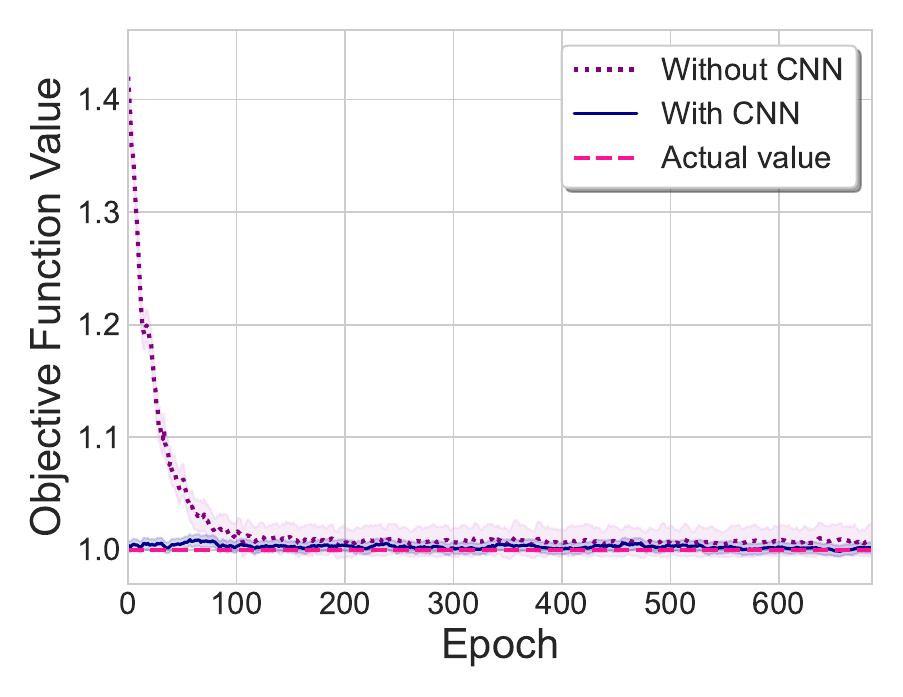}}}

\newcommand\reimagef{\adjustbox{valign=m,vspace=0.1pt}{\includegraphics[width=.30\linewidth]{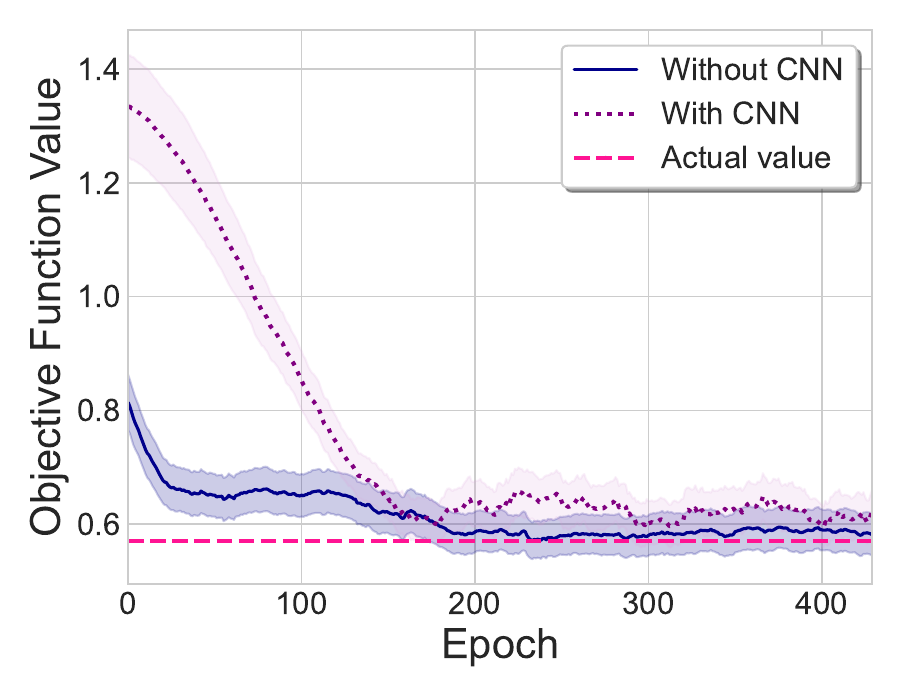}}}
\newcommand\reimages{\adjustbox{valign=m,vspace=0.1pt}{\includegraphics[width=.30\linewidth]{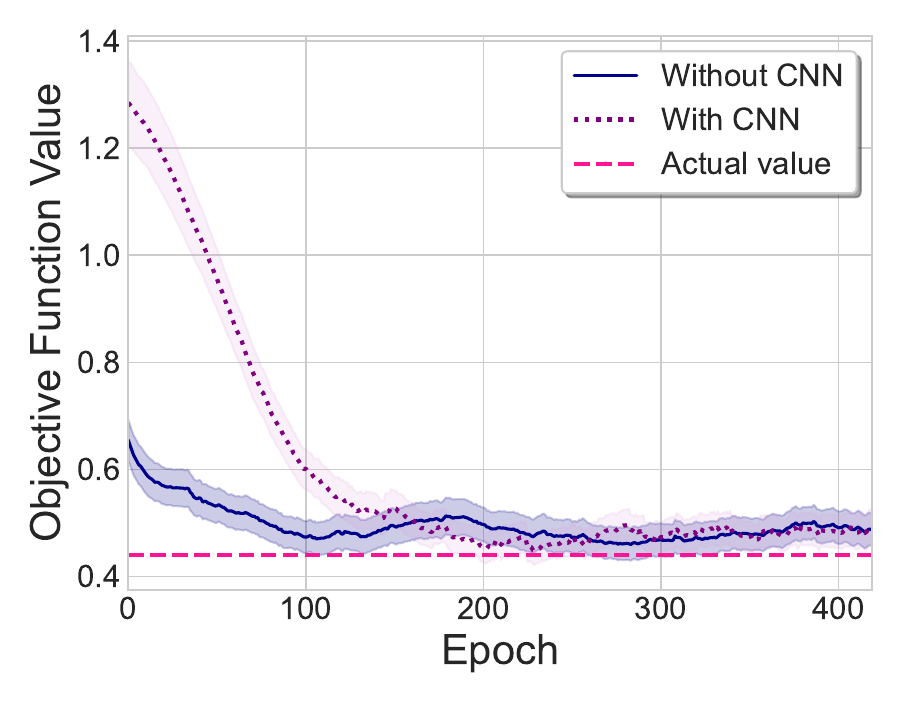}}}

\newcommand\mrreimagef{\adjustbox{valign=m,vspace=0.1pt}{\includegraphics[width=.30\linewidth]{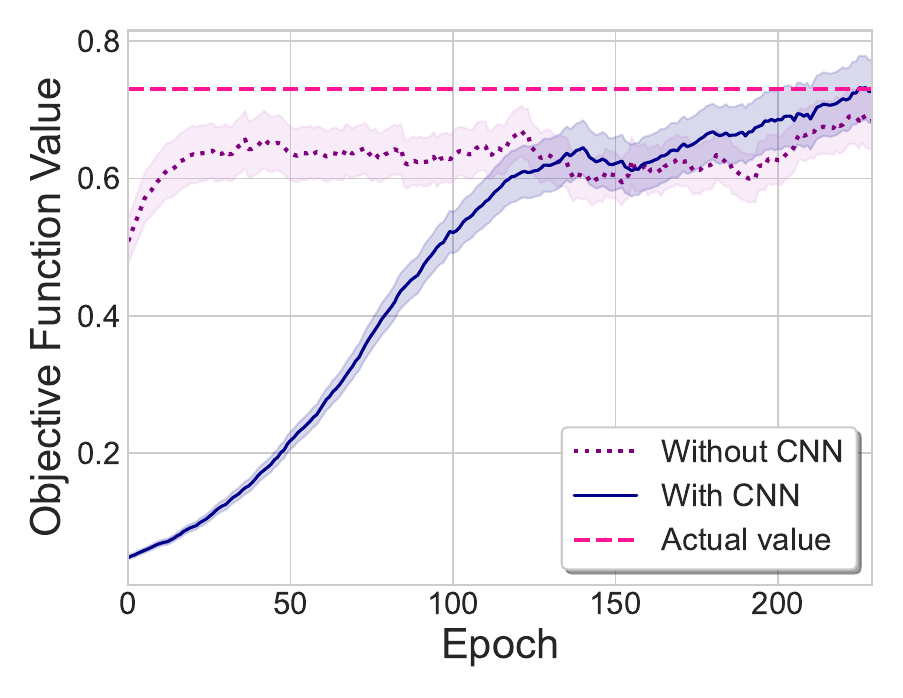}}}
\newcommand\mrreimages{\adjustbox{valign=m,vspace=0.1pt}{\includegraphics[width=.30\linewidth]{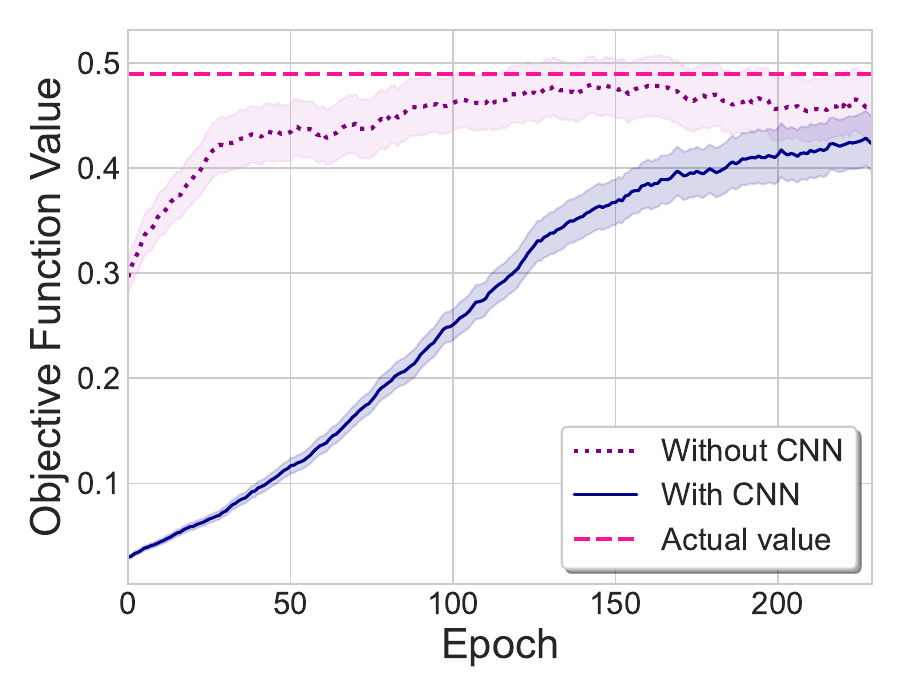}}}

\newcommand\mreimagef{\adjustbox{valign=m,vspace=0.1pt}{\includegraphics[width=.30\linewidth]{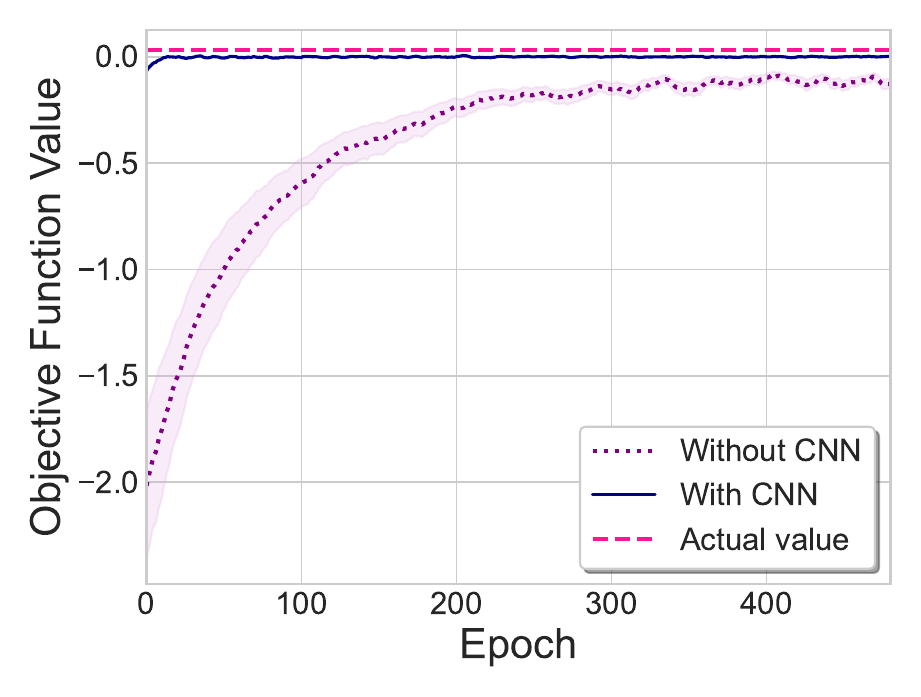}}}
\newcommand\mreimages{\adjustbox{valign=m,vspace=0.1pt}{\includegraphics[width=.30\linewidth]{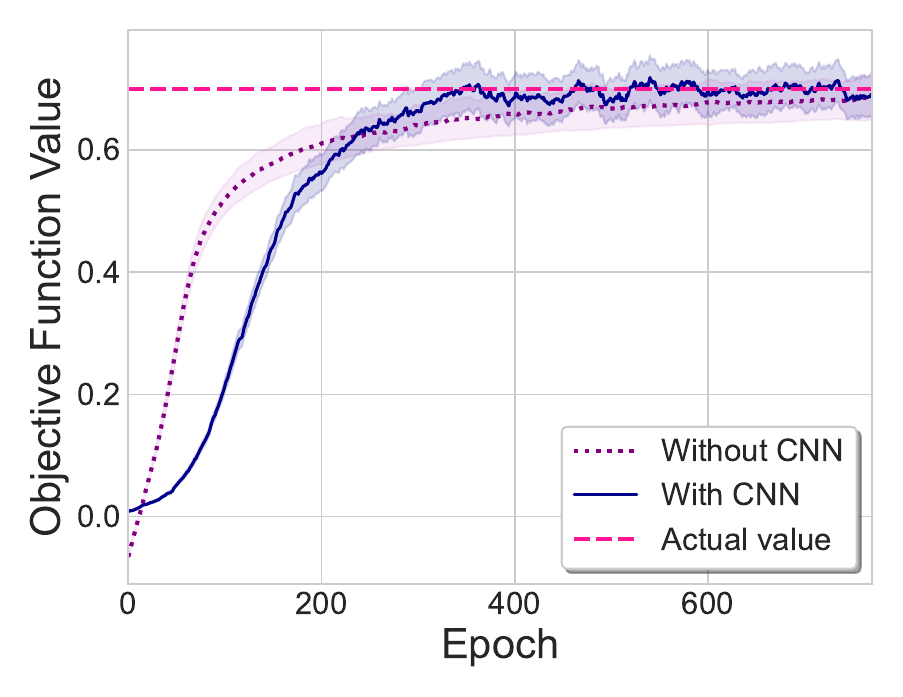}}}

\newcommand\vnimagefsix{\adjustbox{valign=m, vspace=0.1pt}{\includegraphics[width=.30\linewidth]{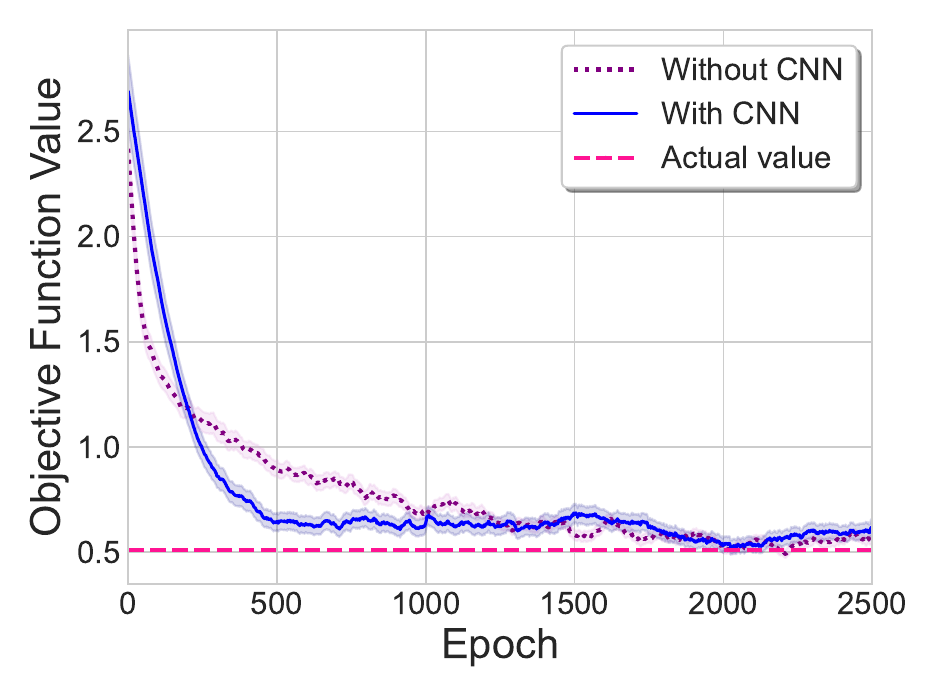}}}
\newcommand\vnimagessix{\adjustbox{valign=m,vspace=0.1pt}{\includegraphics[width=.30\linewidth]{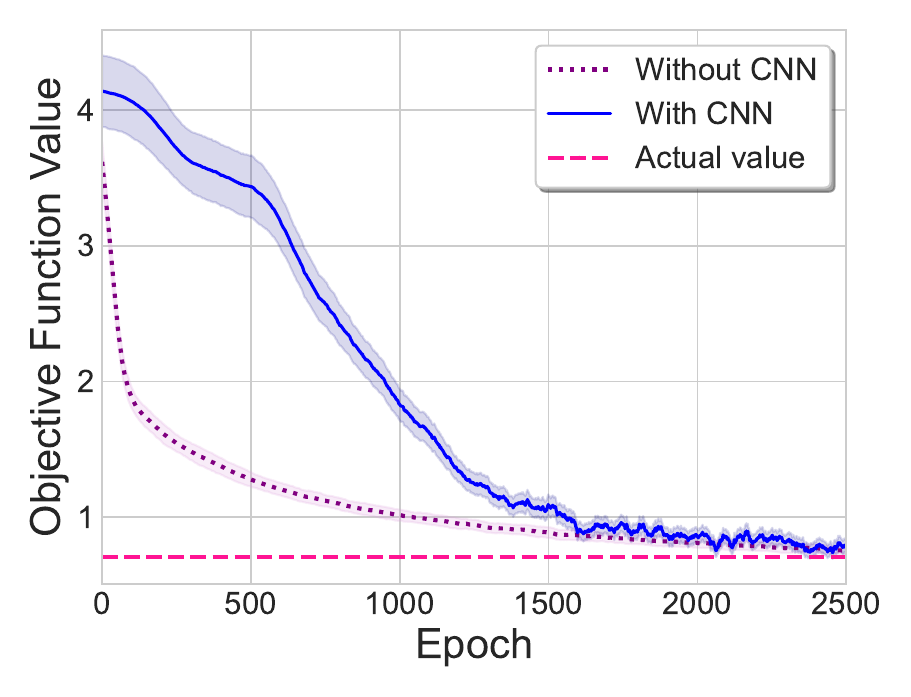}}}

\newcommand\fiimagefsix{\adjustbox{valign=m,vspace=0.1pt}{\includegraphics[width=.30\linewidth]{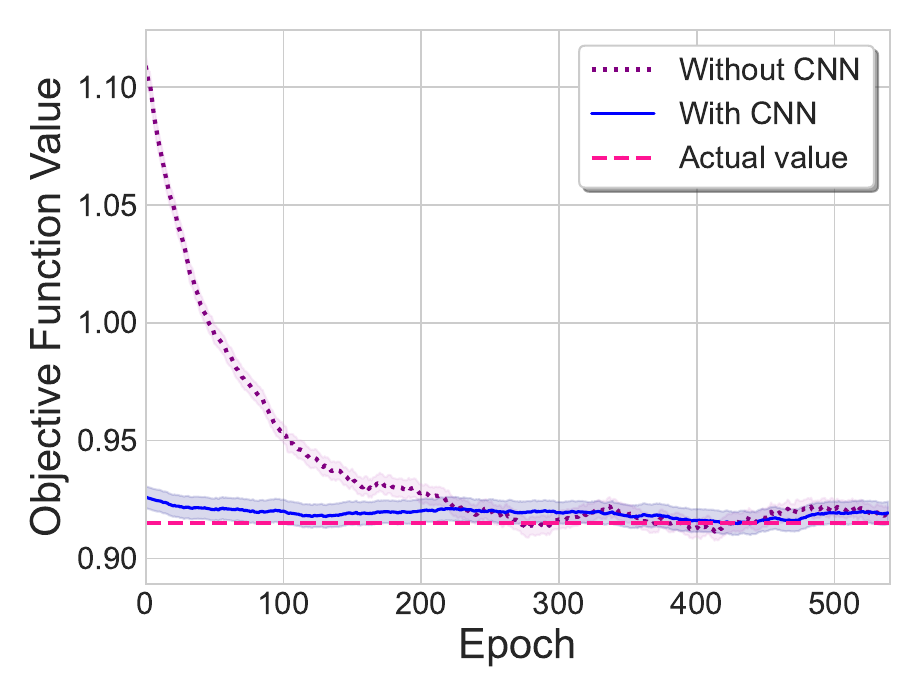}}}
\newcommand\fiimagessix{\adjustbox{valign=m,vspace=0.1pt}{\includegraphics[width=.30\linewidth]{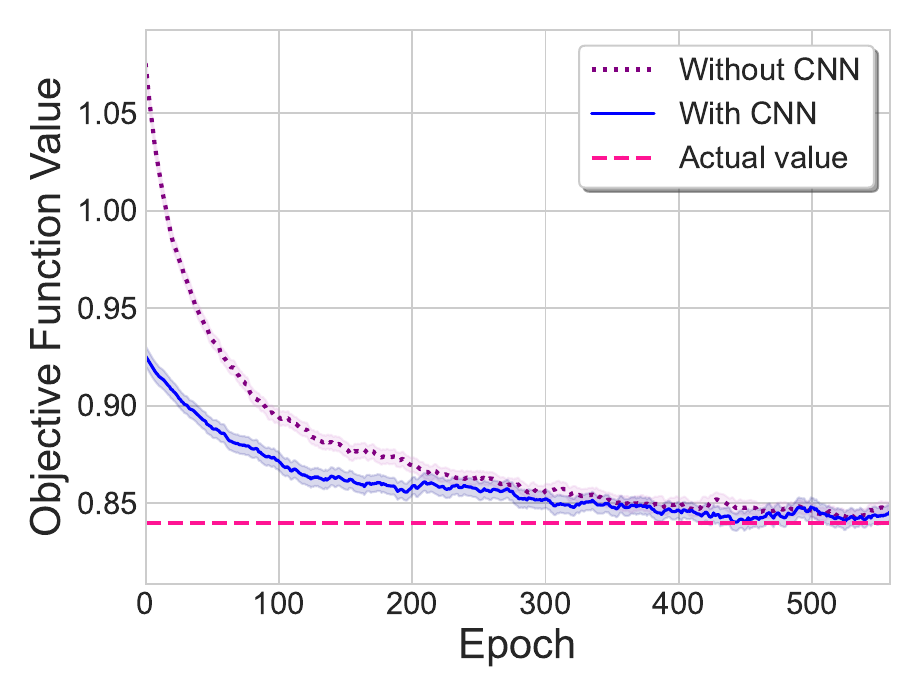}}}

\newcommand\reimagefsix{\adjustbox{valign=m,vspace=0.1pt}{\includegraphics[width=.30\linewidth]{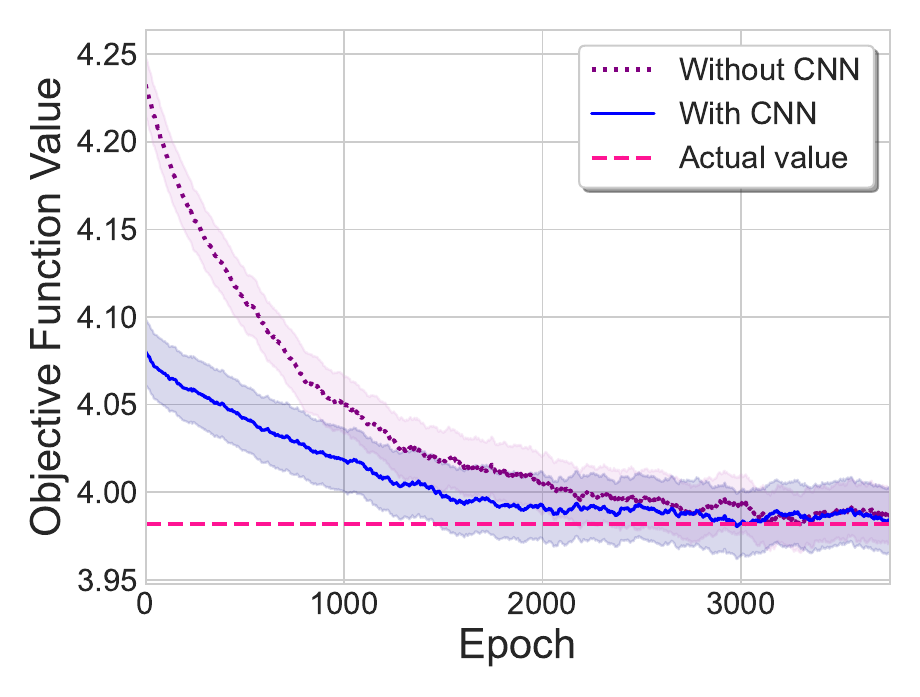}}}
\newcommand\reimagessix{\adjustbox{valign=m,vspace=0.1pt}{\includegraphics[width=.30\linewidth]{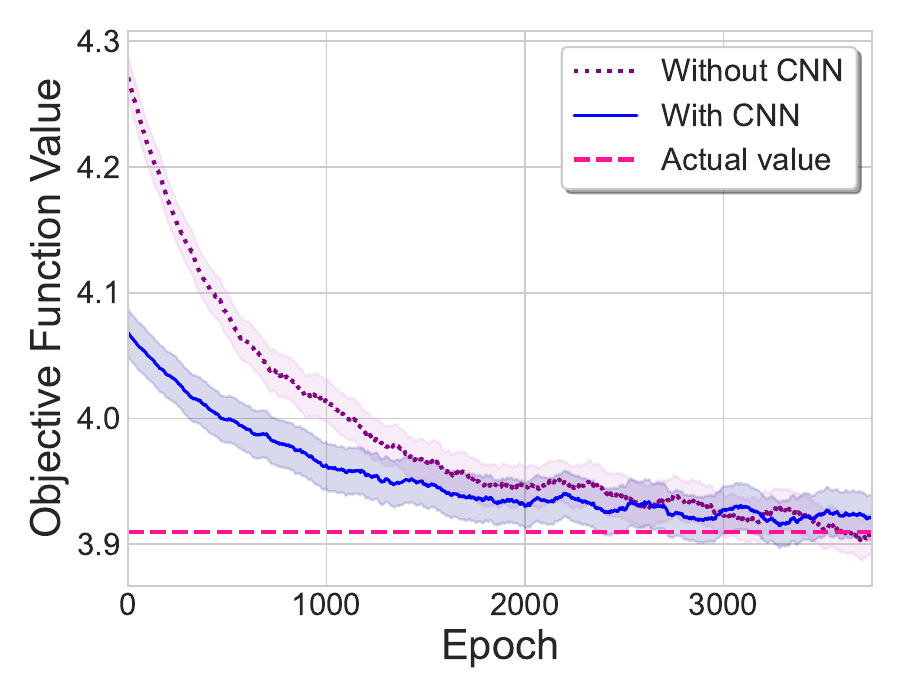}}}

\newcommand\mrreimagefsix{\adjustbox{valign=m,vspace=0.1pt}{\includegraphics[width=.30\linewidth]{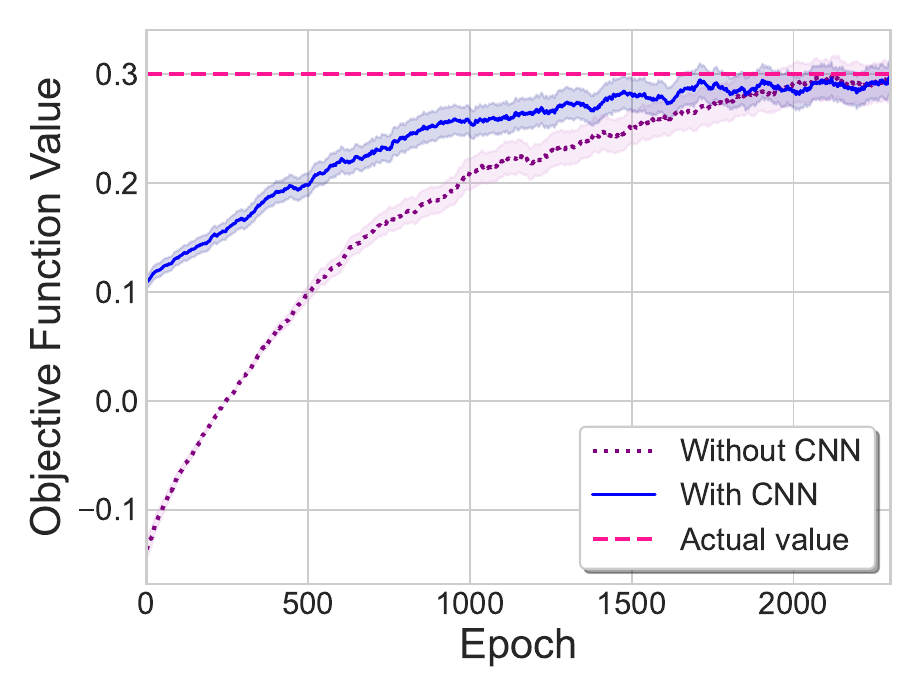}}}
\newcommand\mrreimagessix{\adjustbox{valign=m,vspace=0.1pt}{\includegraphics[width=.30\linewidth]{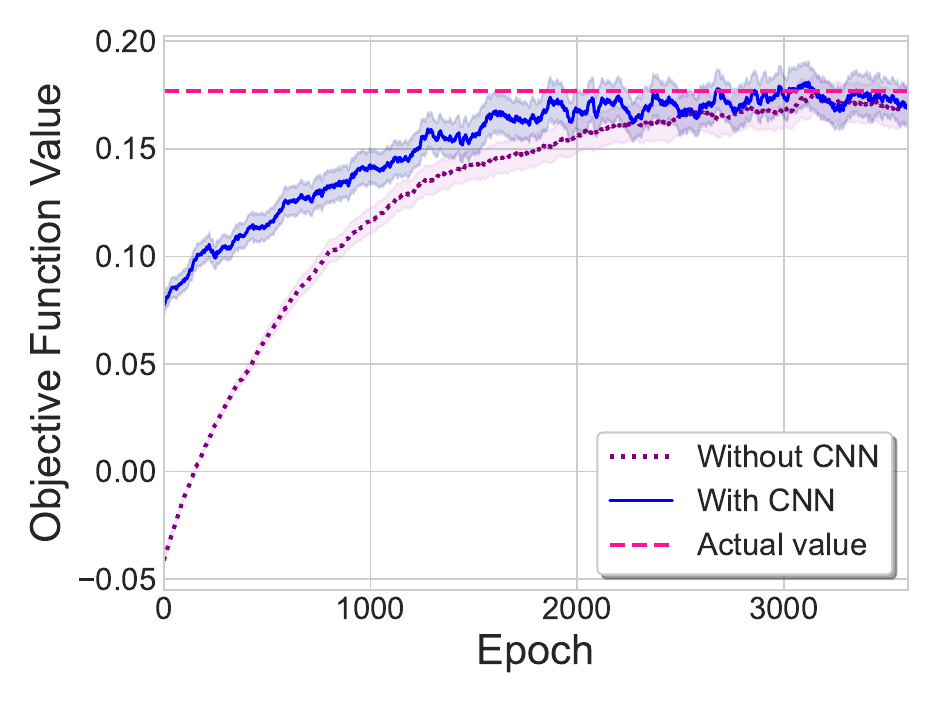}}}

\newcommand\mreimagefsix{\adjustbox{valign=m,vspace=0.1pt}{\includegraphics[width=.30\linewidth]{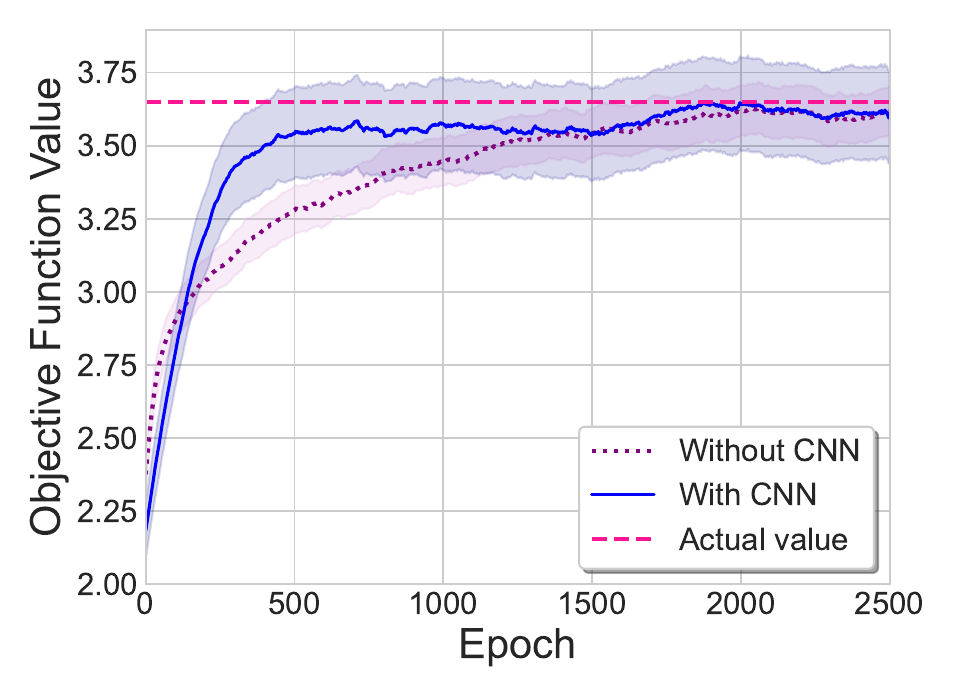}}}
\newcommand\mreimagessix{\adjustbox{valign=m,vspace=0.1pt}{\includegraphics[width=.30\linewidth]{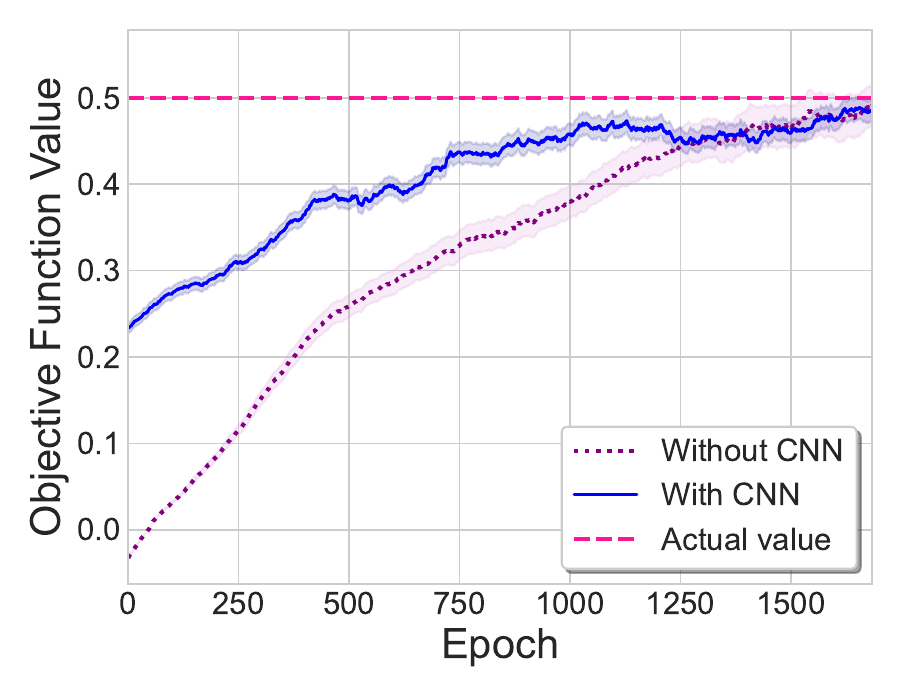}}}

\providecommand{\U}[1]{\protect\rule{.1in}{.1in}}

\newtheorem{remark}{Remark}

\numberwithin{theorem}{section}
\numberwithin{proposition}{section}
\numberwithin{definition}{section}
\numberwithin{example}{section}

\newcommand{\norm}[1]{\left\Vert#1\right\Vert}

\newcommand{\sD}{\mathsf{D}}

\newcommand{\sF}{\mathsf{F}}

\newcommand{\sH}{\mathsf{H}}

\newcommand{\cL}{\mathcal{L}}

\newcommand{\cX}{\mathcal{X}}

\newcommand{\RR}{\mathbb{R}}

\newcommand\numberthis{\addtocounter{equation}{1}\tag{\theequation}}
\allowdisplaybreaks

\begin{document}

\preprint{APS/123-QED}

\title{Quantum Neural Estimation of Entropies}

\author{Ziv Goldfeld}
\affiliation{School of Electrical and Computer Engineering, Cornell
University, Ithaca, NY 14850, USA}

\author{Dhrumil Patel}
\affiliation{Department of Computer Science, Cornell University, Ithaca,  New York, 14850, USA}

\author{Sreejith Sreekumar}
\affiliation{Institute for Quantum Information, RWTH Aachen
University, Aachen, Germany}
\affiliation{School of Electrical and Computer Engineering, Cornell
University, Ithaca, NY 14850, USA}

\author{Mark M. Wilde}
\affiliation{School of Electrical and Computer Engineering, Cornell
University, Ithaca, NY 14850, USA}

\date{\today}

\begin{abstract}
Entropy measures quantify the amount of information and correlation present in a quantum system. In practice, when the quantum state is unknown and only copies thereof are available, one must resort to the estimation of such entropy measures. Here we propose a variational quantum algorithm for estimating the von Neumann and R\'{e}nyi entropies, as well as the measured relative entropy and measured R\'{e}nyi relative entropy. Our approach first parameterizes a variational formula for the measure of interest by a quantum circuit and a classical neural network, and then optimizes the resulting objective over parameter space. Numerical simulations of our quantum algorithm are provided, using a noiseless quantum simulator. The algorithm provides accurate estimates of the various entropy measures for the examples tested, which renders it as a promising approach for usage in downstream tasks.
\end{abstract}

\maketitle

\tableofcontents

\section{Introduction}\label{sec:intro}

Entropy plays a fundamental role in quantifying uncertainty or the level of disorder
in physical systems \cite{RevModPhys.50.221}, with its roots tracing back to thermodynamics and statistical mechanics \cite{NeumannThermodynamikGesamtheiten}. Since its inception, it has been central for studying various fields of physics and beyond, including cosmology, meteorology, physical chemistry, thermodynamics, and information science. 
One of its critical applications is in explaining the inefficiency of heat-work dynamics and the irreversibility of physical systems via the second law of thermodynamics \cite{lieb1998guide}. Moreover, the concept of entropy increasing over time provides a basis for causality and the arrow of time \cite{halliwell1996physical}, which are fundamental to our understanding of the universe.

From an information-theoretic perspective, Shannon developed the concept of entropy to measure the information content of a source \cite{Shannon1948ACommunication}, and  R\'{e}nyi later extended it to a single-parameter family of information measures~\cite{Rrnyi1961OnInformation}. Various discrepancy measures that are intimately connected to entropy were introduced throughout the years in the information theory and mathematical statistics communities, encompassing relative entropy \cite{Kullback1951OnSufficiency} and $f$-divergences \cite{C67,Ali1966,M63}. Quantum counterparts of these measures include the von Neumann entropy~\cite{NeumannThermodynamikGesamtheiten}, quantum R\'{e}nyi entropy, quantum relative entropy \cite{U62}, and various quantum $f$-divergences \cite{Petz1985Quasi-entropiesAlgebra,Petz1986Quasi-entropiesSystems,Mat13,W18opt,hirche2023quantum}. In particular, the entropy of a quantum state quantifies how mixed it is, which is an important concept in quantum mechanics; cf.~\cite{greven2014entropy} for a detailed survey of the role of entropy in physics. 

While the aforementioned measures are essential for determining the amount of information and correlation present in classical and quantum systems, the state of the system is often unknown in practice and is only accessible through sampling. In such situations, one must resort to estimation of the entropy or divergence terms, which is the task at the core of this work. 

%\vspace{-4mm}
\subsection{Contributions}
%\vspace{-3mm}

Here we propose a variational quantum algorithm (VQA) for estimating the von Neumann and R\'{e}nyi entropies of an unknown quantum state (see \eqref{eq:closed-form-vn} and \eqref{eq:quatRenent} for definitions). Our approach also extends to the measured R\'enyi relative entropy between two unknown quantum states, which, in particular, captures the measured relative entropy and fidelity as special cases (see \eqref{eq:renyi_rel_ent}, \eqref{eq:measured-rel-def}, and \eqref{eq:fidelity-closed-form}, respectively). The proposed VQA is developed based on variational formulas for these entropy measures  \cite{petz2007quantum,watrous2012simpler,Berta2015OnEntropies}, which represent them as optimizations of  certain objectives over the space of Hermitian operators. The key idea behind our approach is to parameterize this optimization domain using two models: (i) a classical neural network to parameterize the spectrum (eigenvalues) of a Hermitian operator, and (ii) a quantum circuit to parameterize the eigenvectors.  Having that, we present a sampling procedure for execution on a quantum computer, using which the parameterized objective is approximated by employing independent copies of the unknown quantum states. This results in an optimization problem over the parameter space, which is solvable using classical tools and whose solution yields an estimate of the entropy measure of interest.

Our VQA-based estimation technique, termed \emph{quantum neural estimation}, builds upon the possibility of VQAs realizing quantum speedup \cite{Cerezo2021VariationalAlgorithms,Jaeger2023} and is fully trainable using classical optimizers. Analogous variational methods based on neural network parameterization have seen immense success in recent years for estimating classical entropies and divergences \cite{pmlr-v70-arora17a,pmlr-v80-belghazi18a,Mroueh_Melnyk_Dognin_Ross_Sercu_2021}. The appeal of such neural estimators stems from their scalability to high-dimensional problems and large datasets, as well as their compatibility with modern gradient-based optimization techniques. We expect the proposed quantum neural estimation framework to enjoy similar virtues. Numerical simulations that demonstrate the performance of our method on small-scale examples are provided, revealing accurate estimates and convergence of the training algorithm. We also discuss extensions of our method to quantum relative entropy~\cite{U62} and sandwiched R\'enyi relative entropy \cite{muller2013quantum,wilde2014strong}, surfacing key challenges and outlining potential avenues to overcome them, which are left for future work. 

%\vspace{-4mm}
\subsection{Literature Review}
%\vspace{-3mm}

Estimation of the von Neumann and R\'{e}nyi entropies has attracted significant interest throughout the years. 
A na\"ive approach is to use quantum tomography to estimate the entire density matrix and then evaluate the entropic quantities based on the estimate. 
However, the time complexity of this approach is linear in the dimension of the state (or, equivalently, exponential in the number of qubits) \cite{HHJWY16,wright2016learn,OW16,Yuen2023improvedsample} and is thus untenable. To address this problem, many quantum algorithms for estimating these quantities have been proposed, and their computational complexities have been investigated under different input models \cite{Li2017QuantumEstimation,Subasi2019EntanglementCircuit,Acharya2019MeasuringEntropy, Subramanian2019QuantumStates,Gilyen2019DistributionalWorld,Gur2021SublinearEntropy, Wang2022QuantumEntropies,Wang2022NewDistances,Gilyen2022ImprovedEstimation,Wang2022QuantumSystems,
 Wang2023QuantumEstimation,wang2024timeefficient}.
The authors of \cite{Acharya2019MeasuringEntropy} studied the problem of estimating the von Neumann and R\'{e}nyi entropies, given access to independent copies of the input quantum state. They demonstrated that the sample complexity (i.e., the number of independent copies of the quantum state) of this task grows exponentially with the number of qubits. A number of other papers have proposed methods for estimating integer R\'enyi entropies \cite{Barenco1997,Buhrman2001,Enk2012,Elben2018,Brydges2019,Leone2023}, with limits on the sample complexities also discussed in \cite{Acharya2019MeasuringEntropy}. Entropy estimation under the quantum query model, i.e., where one has access to an oracle that prepares the input quantum state, was explored in~\cite{Subramanian2019QuantumStates,Gur2021SublinearEntropy}. These works established that the query complexity (i.e., the number of times the oracle is queried) for estimating the von Neumann  and R\'{e}nyi entropies is also exponential in the number of qubits.

Variational methods are often used in physics to find approximate solutions to problems that are hard to solve exactly \cite{GRS83}, e.g., computing ground-state energies for quantum systems \cite{CM16}. As such, they underlie various computational methods, including the Hartree--Fock method \cite{Hartree1928TheDiscussion} and the configuration interaction method~\cite{DavidSherrill1999TheApproaches}. 
VQAs are an extension of classical variational methods, and have become a prominent research area in quantum computing \cite{Cerezo2021VariationalAlgorithms,bharti2021noisy}. VQAs are hybrid quantum-classical algorithms that are driven by classical optimizers and only call a quantum subroutine for tasks that are (presumably) hard for a classical machine. These algorithms apply the variational principle to find approximate solutions, by first preparing a quantum trial state and then optimizing its parameters using a classical computer. VQAs have now been applied to a variety of problems, including quantum simulation~\cite{Yuan2019TheorySimulation}, optimization \cite{Peruzzo2014AProcessor, Farhi2014AAlgorithm, Patel2021VariationalProgramming}, and machine learning~\cite{Biamonte2017QuantumLearning}. A VQA is also at the heart of the quantum neural estimation method developed herein.

Neural estimation techniques have been at the forefront of research on classical entropy and divergence estimation. The idea is to parameterize a variational form of the measure of interest by a neural network, approximate expectations by sample means, and optimize the resulting empirical objective over parameter space. The appeal of neural estimators stems from excellent scalability and computational efficiency observed in practice, which is in line with the success of neural nets for language models~\cite{devlin2018bert,brown2020language}, computer vision~\cite{wang2022yolov7}, and generative modeling~\cite{ramesh2022hierarchical,karras2019style,ramesh2021zero,saharia2022photorealistic}. Various neural estimators of Shannon's mutual information \cite{pmlr-v80-belghazi18a,Mroueh_Melnyk_Dognin_Ross_Sercu_2021}, neural net distances \cite{pmlr-v70-arora17a,Zhang-2018}, and classical $f$-divergences \cite{sreekumar2021non,Sreekumar-Goldfeld-2022} have been developed and analyzed for their performance. Neural estimators of quantum entropy measures have not appeared in the literature yet, and we make strides towards harnessing this promising methodology in this work (see also the concurrent work discussed~below).

\medskip
\noindent\textbf{Note on concurrent work.} The independent and concurrent work~\cite{shin2023estimating} appeared on the arXiv a few days before our preprint~\cite{goldfeld2023quantum} was uploaded. It introduced a method for estimating von Neumann entropy reminiscent of ours, but with a few key differences. First, we treat several prominent quantum entropy measures---von Neumann entropy, R\'enyi entropy, measured relative (R\'enyi) entropy, and fidelity---while \cite{shin2023estimating} only accounts for the von Neumann entropy. Second, we parameterize the space of all Hermitian operators using parameterized quantum circuits and classical neural networks, whereas \cite{shin2023estimating} rewrites the variational formula for von Neumann entropy in terms of an optimization over parameterized quantum states and uses only the quantum circuit component. We believe that the incorporation of classical neural networks is crucial for scalable estimation of a broad class of quantum entropy and divergence measures. 

Let us also note that another independent and concurrent work~\cite{lee2023estimating} appeared on the arXiv roughly one month after our preprint~\cite{goldfeld2023quantum} was uploaded. The approach pursued in \cite{lee2023estimating} is similar to ours presented here, employing quantum neural estimation as a combination of classical neural networks and parameterized circuits to estimate von Neumann and R\'enyi entropies. The applications considered in \cite{lee2023estimating} are to classify phases of a Hamiltonian of physical interest, the $XXZ$ chain, and to compare the performance of quantum neural estimation with a variational quantum state eigensolver. 

\section{Quantum Entropies and Divergences}

In this section, we define the quantum entropy and discrepancy measures of interest and present their variational forms, which are subsequently used for the proposed quantum neural estimation method. Throughout, fix $d\in \mathbb{N}$, and let $\rho$ and $\sigma$ be $d$-dimensional quantum states.

%\vspace{-4mm}
\subsection{Measured Relative Entropy and von Neumann Entropy}
%\vspace{-3mm}

The \emph{measured relative entropy} between $\rho$ and $\sigma$ is defined as \cite{Donald1986,P09}
\begin{equation}
\sD_M(\rho\Vert\sigma)\coloneqq\sup_{\cX,\left(  \Lambda_{x}\right)  _{x\in\cX}}
\sum_{x\in\cX}\operatorname{Tr}[\Lambda_{x}\rho]\ln\!\left(  \frac{\operatorname{Tr}
[\Lambda_{x}\rho]}{\operatorname{Tr}[\Lambda_{x}\sigma]}\right),
\label{eq:measured-rel-def}
\end{equation}
where the supremum is over all finite sets $\cX$ and positive operator-valued measures (POVMs) $\left(  \Lambda_{x}\right)_{x\in\cX}$ indexed by~$\cX$.\footnote{Ref.~\cite{Donald1986} defined the quantity with an optimization over just projective measurements, but Ref.~\cite{P09} generalized the definition to include an optimization over all possible measurements.} That is, it is equal to the largest classical relative entropy between the probability distributions $(\operatorname{Tr}
[\Lambda_{x}\rho])_{x \in \mathcal{X}}$ and $(\operatorname{Tr}
[\Lambda_{x}\sigma])_{x \in \mathcal{X}}$ that result from performing a measurement $(\Lambda_{x})_{x \in \mathcal{X}}$ on the respective states $\rho$ and $\sigma$.\footnote{Here, we identify a probability distribution $p$ on $\cX$ with the $|\cX|$-dimensional simplex vector $(p(x))_{x\in\cX}$.} As a consequence of the data-processing inequality for the classical relative entropy, it suffices to perform the optimization in \eqref{eq:measured-rel-def} over rank-one measurements \cite{Berta2015OnEntropies} (i.e., where each $\Lambda_{x}$, for $x\in\cX$, is a rank-one operator). A variational form for $\sD_M(\rho\Vert\sigma)$ was derived in \cite[Eq.~(11.69)]{petz2007quantum} (see also \cite[Lemma~1 and Theorem~2]{Berta2015OnEntropies}), whereby
\begin{align}
\sD_{M}(\rho\Vert\sigma)  & =\sup_{H}\left\{\operatorname{Tr}[H\rho
]-\ln\operatorname{Tr}[e^H\sigma]\right\}
\label{eq:rel_ent_variational_other}
\\
& = \sup_{H}\left\{\operatorname{Tr}[H\rho
]-\operatorname{Tr}[e^H\sigma]\right\} + 1,\label{eq:rel_ent_variational}
\end{align}
where the supremum in \eqref{eq:rel_ent_variational_other}--\eqref{eq:rel_ent_variational} is over all $d \times d$ Hermitian operators (i.e., $H$ such that $H = H^{\dagger}$). Eqs.~\eqref{eq:rel_ent_variational_other}--\eqref{eq:rel_ent_variational} can be understood as a quantum generalization of the Donsker--Varadhan formula for classical relative entropy \cite[Lemma~2.1]{donsker1975asymptotic} (see also \cite[Lemma~1 and Eq.~(8)]{NWJ10}). Throughout our paper and numerical simulations, we use the variational formula in \eqref{eq:rel_ent_variational}.

An important special case of the measured relative entropy is the \emph{von Neumann entropy}
\begin{equation}
\label{eq:closed-form-vn}
\sH(\rho)\coloneqq -\operatorname{Tr}[\rho\ln\rho].
\end{equation}
It is obtained from $\sD_M(\rho\Vert\sigma)$ by setting $\sigma$ to be the maximally mixed state $\pi_{d}\coloneqq  I/d$, where $I$ is the identity operator. Namely, we have
\begin{equation}
\sH(\rho)=\ln d-\sD_M(\rho\Vert\pi_{d}),
\label{eq:von-Neumann-to-measured}
\end{equation}
which follows from \cite[Section~16]{Hein1979} (see also \cite[Theorem~11.1.1]{wilde2017quantum}).
Inserting the variational representation from~\eqref{eq:rel_ent_variational} into \eqref{eq:von-Neumann-to-measured}, the von Neumann entropy is expressed in the following variational form:
\begin{align}
\sH(\rho)  
=\ln d - 1 -  \sup_{H}\left\{\operatorname{Tr}[H\rho]-
\operatorname{Tr}\!\left[e^H\pi_{d}\right]\right\}. \label{eq:varexpvonNeum}
\end{align}

%\vspace{-4mm}
\subsection{Measured R\'enyi Relative Entropy and R\'enyi Entropy}
%\vspace{-3mm}

The \emph{measured R\'{e}nyi relative entropy} of order $\alpha \in (0, 1) \cup (1,\infty)$ between $\rho$ and $\sigma$ is defined as follows \cite[Eqs.~(3.116)--(3.117)]{F96}:
\begin{multline}
\sD_{M,\alpha}(\rho\Vert\sigma) \coloneqq \\
\sup_{\cX,\left(  \Lambda_{x}\right)  _{x\in\cX}
} \frac{1}{\alpha-1}\ln \sum_{x\in\cX} \operatorname{Tr}[\Lambda_{x}\rho]^{\alpha}\operatorname{Tr}[\Lambda_{x}\sigma]^{1-\alpha} ,
\label{eq:renyi_rel_ent}
\end{multline}
where the supremum is over all finite sets $\cX$ and POVMs \ $\left(  \Lambda_{x}\right)_{x\in\cX}$. Akin to the measured relative entropy case, $\sD_{M,\alpha}(\rho\Vert\sigma)$ is equal to the largest classical R\'enyi relative entropy between probability distributions of the form $(\operatorname{Tr}
[\Lambda_{x}\rho])_{x \in \mathcal{X}}$ and $(\operatorname{Tr}
[\Lambda_{x}\sigma])_{x \in \mathcal{X}}$ that result from performing a measurement $(\Lambda_{x})_{x \in \mathcal{X}}$ on $\rho$ and~$\sigma$. As before, it suffices to perform the optimization in \eqref{eq:measured-rel-def} over rank-one measurements. The definition in \eqref{eq:renyi_rel_ent} recovers the measured relative entropy $\sD_M(\rho\Vert\sigma)$ by taking the limit $\alpha\to 1$. The following variational representation was established in \cite[Lemma~3 and Theorem~4]{Berta2015OnEntropies}:
\begin{equation}
\sD_{M,\alpha}(\rho\Vert\sigma)  =\sup_{H}\left\{\frac{\alpha}{\alpha-1}\ln\operatorname{Tr}[e^{\left(  \alpha-1\right)  H}\rho
]-\ln\operatorname{Tr}[e^{\alpha H}\sigma]\right\}.
\label{eq:renyi_rel_ent_variational}
\end{equation}
Eq.~\eqref{eq:renyi_rel_ent_variational} is a quantum generalization of the more recently derived variational formula for classical R\'enyi relative entropy \cite[Theorem~3.1]{Birrell2021} (i.e., the classical case arises by plugging in commuting density operators $\rho$ and $\sigma$).

Instantiating $\sigma$ in \eqref{eq:renyi_rel_ent} as the maximally mixed state gives rise to the \emph{quantum  R\'{e}nyi entropy}
\begin{equation}
\sH_{\alpha}(\rho)\coloneqq \frac{1}{1-\alpha}\ln\operatorname{Tr}[\rho^{\alpha}],
\label{eq:quatRenent}
\end{equation}
which admits the form
\begin{equation}
\sH_{\alpha}(\rho)=\ln d-\sD_{M,\alpha}(\rho\Vert\pi_{d}).
\end{equation}
Leveraging the variational formula in \eqref{eq:renyi_rel_ent_variational}, 
we obtain
\begin{equation}
\sH_{\alpha}(\rho)\mspace{-2mu}=\mspace{-2mu} \ln d\,-\,\sup_H\left\{\frac{\alpha}{\alpha\mspace{-2mu}-\mspace{-2mu}1}\ln
\operatorname{Tr}[e^{\left(\alpha\mspace{-2mu}-\mspace{-2mu}1\right)  H}\rho]-\ln\operatorname{Tr}
\left[  e^{\alpha H}\pi_{d}\right]\right\}\!.\label{eq:varexpquantRen}
\end{equation}

Another interesting special case of the measured R\'enyi relative entropy is obtained for $\alpha = 1/2$:
\begin{align}
\sD_{M,\frac{1}{2}}(\rho\Vert\sigma) & = 
 -\ln \inf_{\cX,\left(  \Lambda_{x}\right)  _{x\in\cX}
}\left[\sum_{x\in\cX} \sqrt{\operatorname{Tr}[\Lambda_{x}\rho]\operatorname{Tr}[\Lambda_{x}\sigma]}\right]^2\notag \\
&  = -\ln \sF(\rho,\sigma),
\label{eq:fidelity-measured}
\end{align}
where the fidelity of $\rho$ and $\sigma$ is defined as \cite{Uhl76}
\begin{equation}
\label{eq:fidelity-closed-form}
    \sF(\rho,\sigma)\coloneqq \left\| \sqrt{\rho} \sqrt{\sigma}\right\|_1^2.
\end{equation}
The equality in \eqref{eq:fidelity-measured} was established in \cite{fuchs1995mathematical}, which indicates that the fidelity of quantum states $\rho$ and $\sigma$ is achieved by a measurement (i.e., minimizing the classical fidelity of the distributions $(\operatorname{Tr}
[\Lambda_{x}\rho])_{x \in \mathcal{X}}$ and $(\operatorname{Tr}
[\Lambda_{x}\sigma])_{x \in \mathcal{X}}$ over all possible measurements). It thus follows from \eqref{eq:renyi_rel_ent_variational} and \eqref{eq:fidelity-measured} that the negative logarithm of the fidelity has the following variational form:
\begin{equation}
-\ln \sF(\rho,\sigma)  =-\inf_{H}\left\{\ln\operatorname{Tr}[e^{-  H}\rho
]+\ln\operatorname{Tr}[e^{ H}\sigma]\right\}.\label{eq:neg-log-fid_variational}
\end{equation}
Alternatively, by making use of the variational form in \cite[Eq.~(20)]{Berta2015OnEntropies}, we find that the root fidelity has the form
\begin{equation}\label{eq:root-fidelity-measured}
    \sqrt{\sF(\rho,\sigma)} = \frac{1}{2}\inf_{H}\left\{\operatorname{Tr}[e^{-  H}\rho
]+\operatorname{Tr}[e^{ H}\sigma]\right\}, 
\end{equation}
which coincides with the expression from \cite[Eq.~(6)]{watrous2012simpler}.

\section{Quantum Neural Estimation of Entropies}

We develop variational estimators for the entropy and measured relative entropy terms defined in the previous section. Our approach assumes access to a black-box procedure for repeatedly preparing the quantum states $\rho$ and $\sigma$. The key idea is to parameterize the set of Hermitian operators in \eqref{eq:varexpquantRen} using a classical neural network and a quantum circuit. The parameterization procedure, sampling step, and the quantum algorithm to optimize the resulting objective are described~next.

%\vspace{-4mm}
\subsection{Measured Relative Entropy}\label{subsec:meas_rel_ent}
%\vspace{-3mm}

\noindent\textbf{Parameterization.}
The spectral theorem implies that any Hermitian operator $H$ can be decomposed as
\begin{equation}
    H = \sum_{i=1}^d\lambda_{i} |\lambda_{i}\rangle\!\langle \lambda_{i}|,
\end{equation}
where $\lambda_1,\ldots,\lambda_d$ are the eigenvalues and $|\lambda_1\rangle,\ldots,|\lambda_d\rangle$ are the corresponding (orthonormal) eigenvectors.

Our first step is to approximate the set of eigenvalues using a classical neural network $f_{\bm{w}}:\{1,\ldots,d\}\to\RR$ with a parameter vector $\bm{w} \in \mathbb{R}^{p}$, where $p\in \mathbb{N}$. The neural network output $f_{\bm{w}}(i)$, for $i\in\{1,\ldots,d\}$, serves as a proxy for the eigenvalue~$\lambda_i$. We keep the architecture of the neural net (viz., nonlinearity, width, depth, etc.) implicit, in order to maintain flexibility of the approach. Next, we approximate the set of eigenvectors using a parameterized quantum circuit $U(\bm{\theta})$ with a parameter vector $\bm{\theta} \in [0, 2\pi]^{q}$, where $q\in \mathbb{N}$. In practice, the total number of neural network and quantum circuit parameters, i.e., $p+q$, should scale like  $O\big(\text{poly}(\log d)\big)$, so that the optimization of the ensuing VQA is efficient with respect to the number of qubits specifying $\rho$ and $\sigma$. See \cite{verdon2019quantum,Liu_2021,Ezzell_2023} for a similar approach for parameterizing the set of mixed quantum states, and \cite[Eq.~(3)]{verdon2019quantum} for a similar approach to parameterizing Hermitian operators.

The above procedure defines a set of parameterized Hermitian operators, specified as
\begin{equation}
H(\bm{w},\bm{\theta})=\sum_{i=1}^d f_{\bm{w}}(i)U(\bm{\theta})|i\rangle\!\langle i|U^{\dag}(\bm{\theta}),\label{eq:Hermitian_param}
\end{equation}
where $(\bm{w},\bm{\theta}) \in \RR^p\times [0,2\pi]^q\!$ and $\{|i\rangle\}_{i=1}^d$ denotes the computational basis. Using this parameterization, we approximate the measured relative entropy from below as follows:
\begin{equation}
\sD_M(\rho\Vert \sigma)\geq \sup_{\bm{w},\bm{\theta}}\left\{\operatorname{Tr}[H(\bm{w},\bm{\theta})\rho
]-\operatorname{Tr}[e^{H(\bm{w},\bm{\theta})}\sigma]\right\}+1,\label{eq:meas_rel_ent_parameterized}
\end{equation}
which follows from \eqref{eq:rel_ent_variational} and because  $\{H(\bm{w},\bm{\theta})\}_{\bm{w},\bm{\theta}}$ is a subset of all $d\times d$ Hermitian operators.

To further simplify the parameterized objective and arrive at a form that lands well for sampling on a quantum computer, define the following probability distributions on $\{1,\ldots,d\}$:
\begin{align}
p_{\bm{\theta}}^\rho(i)  &  \coloneqq \operatorname{Tr}\!\left[|i\rangle\!\langle i|U^{\dag
}(\bm{\theta})\rho U(\bm{\theta})\right] ,\label{eq:dist_p} \\
q_{\bm{\theta}}^\sigma(i)  &  \coloneqq \operatorname{Tr}\!\left[|i\rangle\!\langle i|U^{\dag
}(\bm{\theta})\sigma U(\bm{\theta})\right].\label{eq:dist_q}
\end{align}
Using \eqref{eq:Hermitian_param}, \eqref{eq:dist_p}, and \eqref{eq:dist_q}, we can write the trace terms from the right-hand side of \eqref{eq:meas_rel_ent_parameterized} as
\begin{align}
    \operatorname{Tr}\!\left[H(\bm{w},\bm{\theta})\rho
\right]&=\sum_{i=1}^d p_{\bm{\theta}}^\rho(i)f_{\bm{w}}(i), \label{eq:estterms1}\\
\operatorname{Tr}\!\left[e^{H(\bm{w},\bm{\theta})}\sigma
\right]&=\sum_{i=1}^d q_{\bm{\theta}}^\sigma(i)e^{f_{\bm{w}}(i)},\label{eq:estterms2}
\end{align}
where \eqref{eq:estterms2} follows because 
\begin{equation}
e^{\beta H(\bm{w},\bm{\theta})}=\sum_{i=1}^d e^{\beta f_{\bm{w}}(i)} U(\bm{\theta}) |i\rangle\!\langle
i|U^{\dag}(\bm{\theta}),
\label{eq:exp-form-reduction}
\end{equation}
for all $\beta \in \mathbb{R}$. 
Inserting \eqref{eq:estterms1}--\eqref{eq:estterms2} into the right-hand side of \eqref{eq:meas_rel_ent_parameterized} yields an objective function that is readily estimated using a quantum computer, as described~next.

\medskip
\noindent\textbf{Sampling.} We use a quantum computer to sample from the distributions $p_{\bm{\theta}}^\rho$ and $q_{\bm{\theta}}^\sigma$. As the procedures are similar, we only describe the steps for the former. We prepare the state~$\rho$, act on it with the parameterized unitary $U^{\dag}(\bm{\theta})$, and then measure in the computational basis to obtain a sample from $p_{\bm{\theta}}^\rho$. 
Repeating this process $n$ times for each distribution, we obtain the samples $i_1(\bm{\theta}),\ldots,i_n(\bm{\theta})$ and $j_1(\bm{\theta}),\ldots,j_n(\bm{\theta})$ from $p_{\bm{\theta}}^\rho$ and $q_{\bm{\theta}}^\sigma$, respectively. With that, we approximate the trace values in \eqref{eq:estterms1}--\eqref{eq:estterms2} by sample means, to arrive at the following objective 
\begin{equation}
\cL_n(\bm{w},\bm{\theta})\coloneqq 1+\frac{1}{n}\sum_{\ell=1}^{n}f_{\bm{w}}(i_\ell(\bm{\theta}))-\frac{1}{n}\sum_{\ell=1}^{n}
e^{f_{\bm{w}}(j_\ell(\bm{\theta}))},
\end{equation}
and the resulting estimator is thus
\begin{equation}
    \hat\sD_M^n\coloneq \sup_{\bm{w},\bm{\theta}}\cL_n(\bm{w},\bm{\theta}).\label{eq:estimator}
\end{equation}

\noindent\textbf{Algorithm.} We present a VQA for performing the optimization in \eqref{eq:estimator}.  
The algorithm uses the parameter-shift rule \cite{Li2017,Mitarai2018,Schuld2018EvaluatingHardware} to update the quantum circuit parameters and standard backpropagation \cite{Goodfellow2016DeepLearning} to update the weights of the neural network. The pseudocode of our algorithm is as follows.

\begin{algorithm}[H]
\caption{VQA for Estimating Measured Relative Entropy}
\begin{algorithmic}[1]
\STATE \textbf{Input:} number of iterations $K$, learning rate $\eta$, number of samples $n$, quantum circuits that prepare $\rho$ and $\sigma$.
\vspace{0.7em}
\STATE $\bm{w}^{1} \leftarrow$ Random initialization in $\mathbb{R}^{p}$.
\vspace{0.7em}
\STATE$\bm{\theta}^{1} \leftarrow$ Random initialization in $[0, 2\pi]^{q}$.
\vspace{0.7em}

\FOR{$k \in \{1, 2, \ldots, K\}$}
\vspace{0.7em}
\STATE \hspace*{0.7em}Evaluate $\nabla_{\bm{\theta}} \mathcal{L}_n(\bm{w}^{k}, \bm{\theta}^{k})$ using the parameter-shift rule.
\vspace{0.7em}
\STATE \hspace*{0.7em}$\bm{\theta}^{k+1} \leftarrow \bm{\theta}^{k} + \eta \nabla_{\bm{\theta}} \mathcal{L}_n(\bm{w}^{k}, \bm{\theta}^{k})$
\vspace{0.7em}
\STATE \hspace*{0.7em}Evaluate $\nabla_{\bm{w}} \mathcal{L}_n(\bm{w}^{k}, \bm{\theta}^{k})$ using backpropagation.
\vspace{0.7em}
\STATE \hspace*{0.7em}$\bm{w}^{k+1} \leftarrow \bm{w}^{k} + \eta \nabla_{\bm{w}} \mathcal{L}_n(\bm{w}^{k}, \bm{\theta}^{k})$.
\ENDFOR
\vspace{0.7em}
\STATE \textbf{Output:} $\mathcal{L}_n(\bm{w}^{K+1}, \bm{\theta}^{K+1})$ as an estimate of $\sD_M(\rho\Vert \sigma)$.
\end{algorithmic}\label{algo:quantumrenyi}
\end{algorithm}

%\vspace{-4mm}
\subsection{Von Neumann Entropy}
%\vspace{-3mm}

The von Neumann entropy $H(\rho)$ can be estimated via a similar approach by appealing to \eqref{eq:varexpvonNeum} (in this context, see also \cite[Eq.~(21)]{verdon2019quantum}).  
This leads to the upper bound
\begin{equation}
\sH(\rho)\leq \ln d - 1 - \sup_{\bm{w},\bm{\theta}}\left\{\operatorname{Tr}[H(\bm{w},\bm{\theta})\rho
]-\operatorname{Tr}[e^{H(\bm{w},\bm{\theta})}\pi_d]\right\},
\end{equation}
and we thus take the estimator to be
\begin{equation}
    \hat\sH_n\coloneq \ln d - 1 -\sup_{\bm{w},\bm{\theta}} \left\{\frac{1}{n}\sum_{\ell=1}^{n}f_{\bm{w}}(i_\ell(\bm{\theta}))-\frac{1}{n}\sum_{\ell=1}^{n}
e^{f_{\bm{w}}(\tilde{j}_\ell(\bm{\theta}))}\right\},
\label{eq:von-neu-estimator}
\end{equation}
where $\tilde{j}_\ell(\bm{\theta}),\ldots, \tilde{j}_\ell(\bm{\theta})$ are samples from the distribution $\tilde{q}_{\bm{\theta}}$, with
\begin{equation}
\tilde{q}_{\bm{\theta}}(i)\coloneqq \operatorname{Tr}\!\left[|i\rangle\!\langle i|U^{\dag
}(\bm{\theta})\pi_d U(\bm{\theta})\right] = \frac{1}{d}.
\end{equation}
Given that the distribution $\tilde{q}_{\bm{\theta}}$ is simply the uniform distribution,
 a quantum computer is not required to sample from it, and so the term $\frac{1}{n}\sum_{\ell=1}^{n}
e^{f_{\bm{w}}(\tilde{j}_\ell(\bm{\theta}))}$ in \eqref{eq:von-neu-estimator} can be evaluated exclusively by a classical sampling approach.

%\vspace{-4mm}
\subsection{Measured R\'enyi Relative Entropy}\label{subsec:meas_renyi_rel_ent}
%\vspace{-3mm}

We estimate the measured R\'enyi relative entropy via a similar approach using the variational representation in \eqref{eq:renyi_rel_ent_variational}. A few minor modifications to the steps in Section \ref{subsec:meas_rel_ent} are required, as delineated next. Using the parameterization of Hermitian operators given in \eqref{eq:Hermitian_param}, we obtain the variational lower bound
\begin{align*}
&\sD_{M,\alpha}(\rho\Vert\sigma)\\
&\ \  \geq \sup_{\bm{w},\bm{\theta}} \left\{\frac{\alpha}{\alpha-1}\ln\operatorname{Tr}[e^{\left(  \alpha-1\right)  H(\bm{w},\bm{\theta})}\rho
]-\ln\operatorname{Tr}[e^{\alpha H(\bm{w},\bm{\theta})}\sigma]\right\}.\numberthis
\label{eq:measured-renyi-lower-bnd}
\end{align*}
With the same definitions of the distributions $p_{\bm{\theta}}^\rho$ and $q_{\bm{\theta}}^\sigma$ as in \eqref{eq:dist_p}--\eqref{eq:dist_q} and using \eqref{eq:exp-form-reduction}, we rewrite the trace terms in \eqref{eq:measured-renyi-lower-bnd} as
\begin{align}
\operatorname{Tr}\!\left[e^{(\alpha-1)H(\bm{w},\bm{\theta})}\rho
\right]&=\sum_{i=1}^d p_{\bm{\theta}}^\rho(i)e^{(\alpha-1)f_{\bm{w}}(i)},\label{eq:renyi_estterms1}\\
\operatorname{Tr}\!\left[e^{\alpha H(\bm{w},\bm{\theta})}\sigma
\right]&=\sum_{i=1}^d q_{\bm{\theta}}^\sigma(i)e^{\alpha f_{\bm{w}}(i)}.\label{eq:renyi_estterms2}
\end{align}

Following the same sampling step as in Section~\ref{subsec:meas_rel_ent}, we approximate the expected values in \eqref{eq:renyi_estterms1}--\eqref{eq:renyi_estterms2} by sample means and arrive at the estimator
\begin{equation}
    \hat\sD_{M,\alpha}^n\coloneq \sup_{\bm{w},\bm{\theta}}\cL_\alpha^n(\bm{w},\bm{\theta}),\label{eq:estimator-measured-renyi}
\end{equation}
where
\begin{multline}
\cL_\alpha^n(\bm{w},\bm{\theta})\coloneqq
\frac{\alpha}
{\alpha-1} \ln \frac{1}{n}\sum_{k=1}^{n}e^{(\alpha-1)f_{\bm{w}}(i_k(\bm{\theta}))}\\
-\ln\frac{1}{n}\sum_{k=1}^{n}
e^{\alpha f_{\bm{w}}(j_k(\bm{\theta}))}.
\end{multline}

%\vspace{-4mm}
\subsection{R\'enyi Entropy}
%\vspace{-3mm}

We briefly state the variational estimator for the R\'enyi entropy. From \eqref{eq:varexpquantRen}, along with our parameterization procedure, we obtain
\begin{align*}
\sH_{\alpha}(\rho)\leq \ln d 
-\,\sup_{\bm{w},\bm{\theta}}\Bigg\{\frac{\alpha}{\alpha-1}&\ln
\operatorname{Tr}[e^{\left(\alpha-1\right)  H(\bm{w},\bm{\theta})}\rho]\\
&\ \ -\ln\operatorname{Tr}\!
\left[  e^{\alpha H(\bm{w},\bm{\theta})}\pi_{d}\right]\Bigg\}.\numberthis
\end{align*}
The estimator that results from replacing expectations with sample means is thus
\begin{align*}
    \hat\sH^n_{\alpha}\coloneq \ln d
-\sup_{\bm{w},\bm{\theta}}\Bigg\{\frac{\alpha}
{\alpha-1} &\ln \frac{1}{n}\sum_{k=1}^{n}e^{(\alpha-1)f_{\bm{w}}(i_k(\bm{\theta}))}\\
&\qquad-\ln\frac{1}{n}\sum_{k=1}^{n}e^{\alpha f_{\bm{w}}(\tilde{j}_k(\bm{\theta}))}\Bigg\},\numberthis
\label{eq:renyi-estimator}
\end{align*}
where $\tilde{j}_\ell(\bm{\theta}),\ldots, \tilde{j}_\ell(\bm{\theta})$ are samples from the uniform distribution.
As in the case of estimating the von Neumann entropy, a quantum computer is not required to sample from the uniform distribution, and so the term $\frac{1}{n}\sum_{k=1}^{n}e^{\alpha f_{\bm{w}}(\tilde{j}_k(\bm{\theta}))}$ above can be evaluated using a classical sampling approach.

%\vspace{-4mm}
\subsection{Root Fidelity}
%\vspace{-3mm}

Lastly, we estimate the root fidelity $\sqrt{\sF(\rho,\sigma)}$ via a similar approach using the variational representation in \eqref{eq:root-fidelity-measured}. Employing the same parameterization of Hermitian operators given in \eqref{eq:Hermitian_param}, we obtain the variational upper bound
\begin{equation}
 \sqrt{\sF(\rho,\sigma)}
  \leq \frac{1}{2}\inf_{\bm{w},\bm{\theta}}
\left\{\operatorname{Tr}[e^{-  H(\bm{w},\bm{\theta})}\rho
]+\operatorname{Tr}[e^{ H(\bm{w},\bm{\theta})}\sigma]\right\},
\label{eq:fidelity-upper-bnd}   
\end{equation}
and then rewrite the trace terms as 
\begin{align}
\operatorname{Tr}\!\left[e^{-H(\bm{w},\bm{\theta})}\rho
\right]&=\sum_{i=1}^d p_{\bm{\theta}}^\rho(i)e^{-f_{\bm{w}}(i)},\label{eq:fid_estterms1}\\
\operatorname{Tr}\!\left[e^{ H(\bm{w},\bm{\theta})}\sigma
\right]&=\sum_{i=1}^d q_{\bm{\theta}}^\sigma(i)e^{ f_{\bm{w}}(i)},\label{eq:fid_estterms2}
\end{align}
where $p_{\bm{\theta}}^\rho$ and $q_{\bm{\theta}}^\sigma$ are given in \eqref{eq:dist_p}--\eqref{eq:dist_q}.

Following the same sampling step as in Section~\ref{subsec:meas_rel_ent}, we approximate the expected values in \eqref{eq:fid_estterms1}--\eqref{eq:fid_estterms2} by sample means and arrive at the estimator for the root fidelity
\begin{equation}
    \hat{\sF}_n\coloneq \inf_{\bm{w},\bm{\theta}}\cL_F^n(\bm{w},\bm{\theta}),\label{eq:estimator-fid}
\end{equation}
where
\begin{equation}
\cL_F^n(\bm{w},\bm{\theta})\coloneqq
\frac{1}
{2n}  \sum_{k=1}^{n}\left(e^{-f_{\bm{w}}(i_k(\bm{\theta}))}
+
e^{ f_{\bm{w}}(j_k(\bm{\theta}))}\right).
\label{eq:fid-estimator-unoptimized}
\end{equation}

\begin{remark}[Comparison to existing variational estimators]
Alternative variational methods for estimating fidelity have been proposed recently \cite{chen2021variational,rethinasamy2021estimating}. Some of the methods rely on Uhlmann's theorem \cite{Uhl76}, which involves a maximization. However, those approaches require purifications of the states $\rho$ and $\sigma$ in order to estimate their fidelity. Since state purifications are not easily attainable from samples, the fact that our algorithm avoids this need 
represents an advantage. Another variational method was proposed in \cite[Algorithm~7]{rethinasamy2021estimating}. While this algorithm does not require purifications, the estimator employed there for the objective function is biased. 
Our approach, on the other hand, estimates the  
objective function from the right-hand side of \eqref{eq:fidelity-upper-bnd} in an unbiased fashion, as given in \eqref{eq:fid-estimator-unoptimized}. 

\end{remark}

\section{Implementation and Experiments}

We numerically simulate our quantum neural estimation algorithm and assess its accuracy for estimating  
the measured relative entropy, von Neumann entropy,  measured R\'enyi relative entropy, and R\'enyi entropy. 
For each quantity, we benchmark the performance against an estimator that 
avoids the use of a classical neural network and instead explicitly stores the eigenvalues. 
The latter approach is more expressive than the former; however, storing all eigenvalues is infeasible for systems with a large number of qubits. The quantum neural estimation approach circumvents this overhead by parameterizing the eigenvalues with a reasonably-sized classical neural network. The comparison between the two approaches aims to surface the potential performance loss that results for the neural net parameterization, but no noticeable drop in performance is seen in our simulations. 
For our simulations, we use Pennylane's ``default-qubit'' as our quantum circuit simulator~\cite{bergholm2022pennylane}. We next describe our methodology and then follow with the results.

%\vspace{-4mm}
\subsection{Methodology}
%\vspace{-3mm}

All our simulations follow the approach described below.

\medskip
\noindent\textbf{Preparing input quantum states.} We consider examples for which the input states are either two-qubit or six-qubit mixed states. This is to give a sense of how the algorithms scale with the number of qubits. 
To prepare these input mixed states, we first prepare purifications thereof and then trace out the reference systems. Formally, let $\rho_S$ be a mixed state of a system $S$, and let $| \psi \rangle\!\langle \psi|_{RS}$ be a purification of $\rho_{S}$, such that $\operatorname{Tr}_{R}\!\left[ |\psi\rangle\!\langle\psi|_{RS}\right] = \rho_{S}$, where $R$ is a reference system. For our purposes, it is sufficient to consider the reference system to have the same dimension as the system.
That is, in order to prepare a six-qubit mixed state, we prepare its 12-qubit purification. 
Overall, with this approach, we generate four input instances (a pair of two-qubit mixed states and a pair of six-qubit mixed states) for each quantity of interest.

\medskip
\noindent\textbf{Parameterized quantum circuits.}  Our algorithm samples from certain distributions by applying a quantum circuit $U^{\dag
}(\bm{\theta})$ to the input state (cf.,~e.g.,  \eqref{eq:dist_p}). 
We use Pennylane's \texttt{qml.RandomLayers} subroutine to prepare a parameterized quantum circuit with a random structure and then keep this structure fixed throughout multiple runs of a specific simulation. 
We only change the structure when it is not sufficiently expressive, in the sense that the set of generated unitaries is not comparable to the set of all unitaries. The \texttt{qml.RandomLayers} subroutine creates a parameterized quantum circuit with multiple layers, each built by randomly selecting a subset of qubits and applying single-qubit or two-qubit parameterized quantum gates to them. For a two-qubit example, we use three layers, each with three to four parameterized quantum gates, for a total of 9 to 12 parameters. For a six-qubit example, we use five layers, each with three to four parameterized quantum gates, for a total of 15 to 20 parameters. To evaluate the gradient of a given cost function with respect to these parameters, we use the parameter-shift rule~\cite{Li2017,Mitarai2018,Schuld2018EvaluatingHardware}.

\medskip
\noindent\textbf{Classical neural network.} The neural estimator uses a classical neural network to parameterize eigenvalues of Hermitian observables. For our simulations, we consider a 2--10--1 fully connected architecture with sigmoidal activations for two-qubit examples and a 6--30--1 fully connected architecture with sigmoidal activations for six-qubit examples. Here, 2 and 1 are the input and output dimensions, respectively, for the two-qubit examples, and similarly, 6 and 1 are the input and output dimensions for the six-qubit examples. 
A sigmoid is not applied at the output since it restricts the values to [0,1], while the eigenvalues that the neural networks aim to approximate may be outside this interval. 
Gradients are evaluated using the PyTorch automatic differentiation subroutine  \texttt{torch.autograd}.

\medskip
\noindent\textbf{Number of runs.} We estimate expectations of the form $\operatorname{Tr}[\cdot]$, which appear in the objective functions of our quantities, using sample means 
(cf.,~e.g., \eqref{eq:estterms1} or \eqref{eq:estterms2}). 
We find that 100 samples suffice for our experiments, and so we use that sample size throughout. 
We plot the mean (solid or dotted line) and standard deviation (shaded area) of each experiment over 10 runs. 

%\vspace{-3mm}
\subsection{Results}
\begin{figure*}
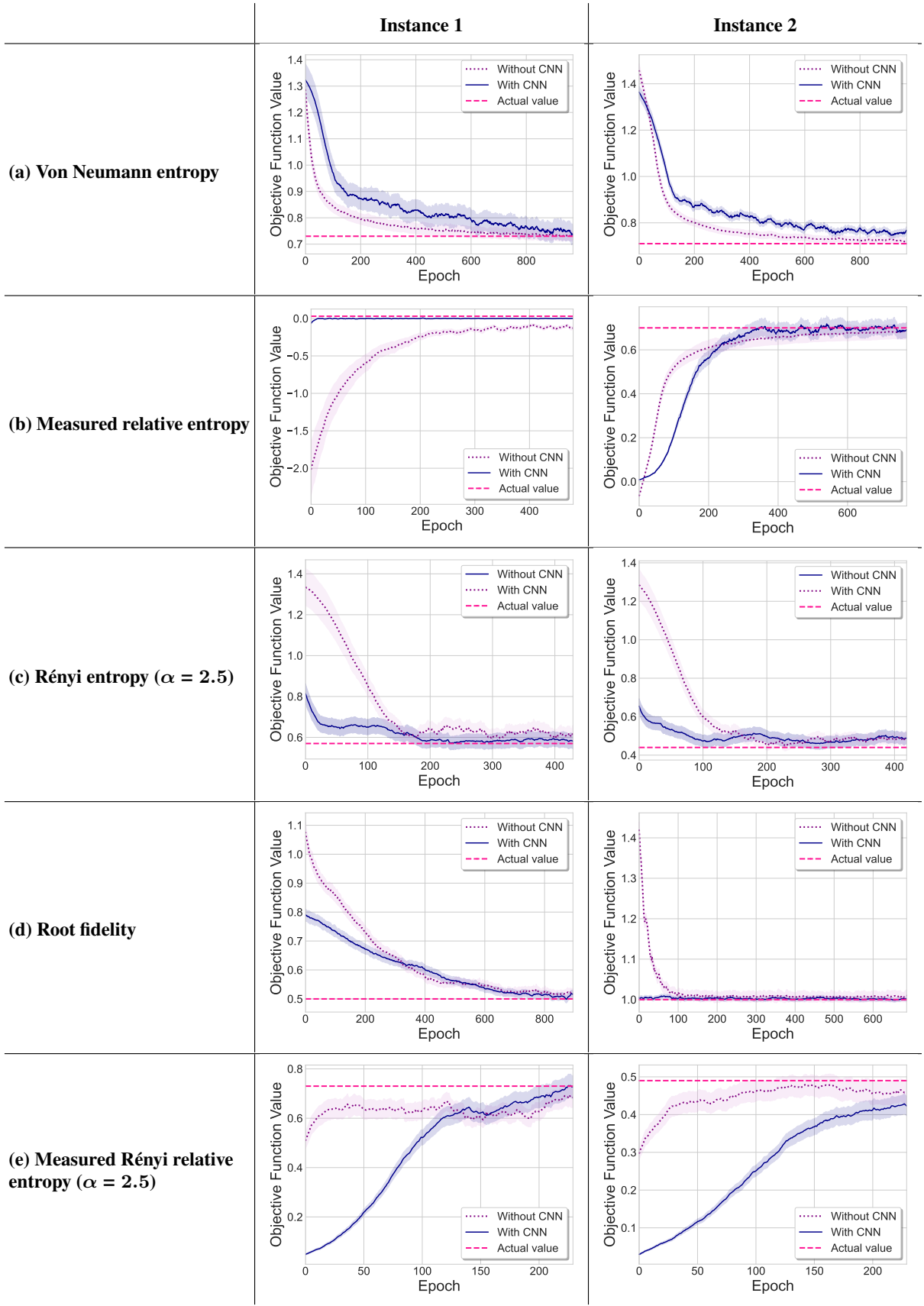

\renewcommand{\arraystretch}{1.8}
\centering
\resizebox{.9\linewidth}{!}{
\begin{tabular}{m{4cm}|c|c}
% \hline
       & \textbf{Instance 1} & \textbf{Instance 2} \\
       \hline
\raggedright \textbf{(a) Von Neumann entropy}  & \vnimagef & \vnimages \\
\hline
\raggedright\textbf{(b) Measured relative entropy}  & \mreimagef & \mreimages \\
\hline
\raggedright\textbf{(c) R\'enyi entropy ($\bm{\alpha = 2.5}$)} & \reimagef & \reimages \\
\hline
\raggedright\textbf{(d) Root fidelity}  & \fiimagef & \fiimages \\
\hline
\raggedright \textbf{(e) Measured R\'enyi relative entropy ($\bm{\alpha = 2.5}$)}  & \mrreimagef & \mrreimages
% \hline
\end{tabular}
}
\caption{ Two-qubit examples.  Convergence  of the quantum neural estimation algorithm for the von Neumann entropy, measured relative entropy, R\'enyi entropy ($\alpha = 2.5$), root fidelity, and measured R\'enyi relative entropy ($\alpha = 2.5$). Two input instances are generated at random for each of these quantities. For each instance, results are presented both with (blue) and without (purple) the classical neural network approximation of eigenvalues. Each solid or dotted line shows the mean value of the estimate, the shaded region represents the standard deviation, while the ground truth is marked by the dashed line. Python source files for reproducing these plots are available with the arXiv posting of our paper.}
\label{fig:plots}
\end{figure*}

%\vspace{-2mm}
The simulation results for two-qubit and six-qubit examples are shown in Figures~\ref{fig:plots} and \ref{fig:plots_six_qubits}, respectively.

\medskip
\noindent\textbf{Von Neumann entropy.}
Figures~\ref{fig:plots}(a) and \ref{fig:plots_six_qubits}(a)  show that our estimator, both with and without the neural network, accurately retrieves the von Neumann entropy with only a small error. 
We note, however, that the convergence rate is quite slow in this case, as it takes around 600 and 2000 epochs to come within 1\% error of the ground truth for two-qubit and six-qubit examples, respectively. To evaluate the ground truth, we use the closed-form expression of the von Neumann entropy in~\eqref{eq:closed-form-vn}.

\begin{figure*}
\renewcommand{\arraystretch}{1.8}
\centering
\resizebox{.9\linewidth}{!}{
\begin{tabular}{m{4cm}|c|c}
% \hline
       & \textbf{Instance 1} & \textbf{Instance 2} \\
       \hline
\raggedright \textbf{(a) Von Neumann entropy}  & \vnimagefsix & \vnimagessix \\
\hline
\raggedright\textbf{(b) Measured relative entropy}  & \mreimagefsix & \mreimagessix \\
\hline
\raggedright\textbf{(c) R\'enyi entropy ($\bm{\alpha = 1.05}$)} & \reimagefsix & \reimagessix \\
\hline
\raggedright\textbf{(d) Root fidelity}  & \fiimagefsix & \fiimagessix \\
\hline
\raggedright \textbf{(e) Measured R\'enyi relative entropy ($\bm{\alpha = 1.05}$)}  & \mrreimagefsix & \mrreimagessix
% \hline
\end{tabular}
}
\caption{Six-qubit examples. Convergence  of the quantum neural estimation algorithm for the von Neumann entropy, measured relative entropy, R\'enyi entropy ($\alpha = 1.05$), root fidelity, and measured R\'enyi relative entropy ($\alpha = 1.05$). Two input instances are generated at random for each of these quantities. For each instance, results are presented both with (blue) and without (purple) the classical neural network approximation of eigenvalues. Each solid or dotted line shows the mean value of the estimate, the shaded region represents the standard deviation, while the ground truth is marked by the dashed line. Python source files for reproducing these plots are available with the arXiv posting of our paper.}
\label{fig:plots_six_qubits}
\end{figure*}

\medskip

\noindent\textbf{Measured relative entropy.} Figures~\ref{fig:plots}(b) and \ref{fig:plots_six_qubits}(b) 
 plot the quantum neural estimation error curves for the case of measured relative entropy.  To compute the ground truth, we use the fact that the measured relative entropy is equal to the quantum relative entropy when the quantum states are positive definite and commute with each other (see \cite[Prop.~5]{Berta2015OnEntropies}). This perspective allows us to work around the issue that there is no closed-form expression for the measured relative entropy, and use that for quantum relative entropy instead:
\begin{align}
\sD(\rho \| \sigma) & \coloneqq \operatorname{Tr}[\rho (\ln \rho - \ln \sigma)]. \label{eq:quantrelent}
\end{align}

\noindent\textbf{R\'enyi entropy.} Figures~\ref{fig:plots}(c) and \ref{fig:plots_six_qubits}(c) present simulations for the R\'enyi entropy of order $\alpha = 2.5$ and $\alpha = 1.05$, respectively. The closed-form expression from \eqref{eq:quatRenent} is used to compute the ground truth. Again, we observe that the neural estimator converges fast and accurately recovers the true R\'enyi entropy value. 

\medskip
\noindent\textbf{Root fidelity.} We simulate quantum neural estimation of  the root fidelity (see \eqref{eq:root-fidelity-measured}) and show the results in Figures~\ref{fig:plots}(d) and~\ref{fig:plots_six_qubits}(d). 
The ground truth fidelity value is computed from~\eqref{eq:fidelity-closed-form}.
We see that the model that employs the neural network approaches the ground truth faster than the one that directly optimizes the eigenvalues. For the second instance shown in Figure~\ref{fig:plots}(d), the neural estimator fluctuates around the ground truth from the onset, while the estimator without the neural net approaches it from above. This effect persists in repeated runs of the simulation.

\medskip

\noindent\textbf{Measured R\'enyi relative entropy.} Figures~\ref{fig:plots}(e) and \ref{fig:plots_six_qubits}(e) plot the quantum neural estimation error curves for the case of measured R\'enyi relative entropy with $\alpha=2.5$ and $\alpha=1.05$, respectively.  To compute the ground truth, we use the fact that the measured R\'enyi relative entropy is equal to the sandwiched R\'enyi relative entropy when the quantum states are positive definite and commute with each other (see \cite[Thm.~6]{Berta2015OnEntropies}). This perspective allows us to get around the fact that there is no closed-form expression for measured R\'enyi relative entropy and instead use that of sandwiched R\'enyi relative entropy:
\begin{align}
\widetilde{\sD}_{\alpha}(\rho \| \sigma) & \coloneqq \frac{1}{\alpha-1}\ln \operatorname{Tr}\left[\left(\sigma^{\frac{1-\alpha}{2\alpha}}\rho \sigma^{\frac{1-\alpha}{2\alpha}}\right)^{\alpha}\right]. \label{eq:sand-renyi-relent}
\end{align}

\section{Concluding Remarks and Future Work}

This work proposed a quantum neural estimation algorithm for estimating quantum information measures, spanning various entropies and measured relative entropies. Our estimator utilizes parameterized quantum circuits and classical neural networks to approximate a variational form of the measure of interest, which enables efficient estimation by sampling and optimization. Numerical experiments that validate the accuracy of the proposed approach on two-qubit and six-qubit settings were provided. Scalability of this approach to larger instances and extensions to other quantum entropies and divergences are important future avenues that we briefly discuss below. It is also of interest to obtain formal performance guarantees for our VQA in terms of its run-time and copy complexity, i.e., the number of quantum states required for the algorithm to achieve a given accuracy.

\medskip
\noindent\textbf{Scalability and the barren-plateau problem.} To achieve the desired scalability, one would have to overcome the barren-plateau problem \cite{McClean_2018}. It refers to the phenomenon that the gradients of a cost function with respect to the quantum circuit parameters in VQAs tend to become exponentially small as the circuit depth and the number of qubits grow. This can have a significant impact on the performance and scalability of VQAs, limiting their ability to find optimal solutions efficiently. Addressing the barren-plateau problem is an active area of research, and various  heuristic approaches towards mitigating its impact have been proposed \cite{Kiani2020,Patti_2021,Liu_2023,rudolph2022synergy,liu2022mitigating}. In particular, we think that the approach of \cite{rudolph2022synergy} is quite promising for overcoming the barren-plateau problem. The authors of \cite{rudolph2022synergy} considered  the variational quantum eigensolver problem and  searched for an initial state  by means of a classical tensor-network approach. They then used this pretrained state as an initialization for a variational quantum algorithm. These authors gave numerical evidence that this approach avoids the barren-plateau problem for examples consisting of up to 100 qubits. It is an important open question to determine how to apply the method of \cite{rudolph2022synergy} to quantum neural estimation of entropies. The main difference between our setting and that of  \cite{rudolph2022synergy} is that the states $\rho$ and $\sigma$ are provided in sample form on a quantum computer, and so it is not clear how to simulate the measurement of observable expectations classically and efficiently via tensor networks.

In the current paper, we did not explicitly investigate the presence of barren plateaus in the context of our quantum neural estimation algorithm. Even though our numerical simulations show good performance, they do not involve sufficiently many qubits for observing barren plateaus. We speculate that the presence of a classical neural network may help to alleviate the barren-plateau problem. This is due to the fact that the cost functions considered in this work are dependent not only on the quantum circuit parameters but also on the neural network parameters. As a result, while gradients with respect to quantum circuit parameters may vanish exponentially with the number of qubits, this may not be the case for gradients with respect to neural network parameters (which are independent of the number of qubits and the type of parameterized quantum circuit used). We leave for future work an in-depth study of this possibility and, generally, of the optimization landscapes. In addition, we plan to explore potential restrictions and formal limitations that may result from barren plateaus and design regularization methods to mitigate them. A promising method towards that end is to first use a tensor network to find a good initialization point for the quantum circuit parameters, and then run the VQA. As mentioned above, this approach was recently shown to provide advantage for the variational quantum eigensolver problem in terms of alleviating barren plateaus and improving the algorithm run-time~\cite{rudolph2022synergy}.

\medskip
\noindent\textbf{Full-state tomography and performance guarantees.} 
An appealing research avenue is to gauge how efficient quantum neural estimation is in comparison to an algorithm that estimates the state via tomography and then computes entropy based on the estimate. When the dimension of the states is small, as in the numerical examples presented herein, we anticipate that full-state tomography is more efficient than quantum neural estimation since the former does not require an optimization over classical or quantum circuit parameters. However, we expect that when the dimension is moderate to large, quantum neural estimation will be more efficient than tomography-based approaches. A formal comparison would require establishing the copy complexity of quantum neural estimation along with its associated run-time. A plausible approach for obtaining an upper bound on this quantity is to first quantify the various components of the overall error, such as approximation, estimation, and optimization errors, in terms of the number of the copies of state used and the circuit parameters. Then the parameters can be optimized to minimize the error of the algorithm and its run-time. We leave this aspect as an open question for future investigation.  

\medskip
\noindent\textbf{Choice of Hamiltonian ansatz.}
The parameterization  in \eqref{eq:Hermitian_param} accounts for a general Hamiltonian ansatz. Namely, the classical neural net and the quantum circuit parametrize the eigenvalues and eigenvectors, respectively, of the class of all Hermitian operators (Hamiltonians), without restricting the structure or complexity of these circuits. However, provided additional knowledge on the underlying quantum states, one may restrict the optimization to a smaller class of Hamiltonians without sacrificing performance, while possibly reducing computational or statistical complexity. For instance, if the eigenvalues or eigenvectors of considered states are from a bounded set or a restricted class of unitaries, then one may consider an appropriate ansatz that encodes this structure into the classical neural network and parameterized quantum circuit. Constraints may also arise organically from the considered quantum divergence. A pertinent example is the normalized trace distance $\frac{1}{2} \norm{\rho-\sigma}_1= \sup_{0 \leq M \leq I}\operatorname{Tr}[M(\rho-\sigma)]$, in which case one may restrict the output of the classical neural net parametrizing $M$ to $[0,1]$. This could potentially lead to computational and/or statistical gains, as is often observed in the classical setting.

\medskip
\noindent\textbf{Other quantum relative entropies.} Another important direction is to extend the quantum neural estimation approach to the relative entropy \cite{U62} and the sandwiched R\'enyi relative entropy \cite{muller2013quantum,wilde2014strong}, defined respectively, in \eqref{eq:quantrelent} and \eqref{eq:sand-renyi-relent}.
These quantities admit the following variational forms (see \cite{Petz1988,kosaki1986relative,Berta2015OnEntropies} and \cite{FL13,Berta2015OnEntropies}, respectively):
\begin{align}
\sD(\rho \| \sigma) & = \sup_{H} \left\{\operatorname{Tr}[H\rho]-\ln\operatorname{Tr}
[\exp\left( H + \ln\sigma\right)  ]\right\},
\label{eq:umegaki-variational-other}\\
& = \sup_{H} \left\{\operatorname{Tr}[H\rho]-\operatorname{Tr}
[\exp\left( H + \ln\sigma\right)  ]\right\}+1,
\label{eq:umegaki-variational}\\
 \widetilde{\sD}_{\alpha}(\rho\Vert\sigma) & = \sup_{H}\mspace{-2mu}\left\{\frac{\alpha}
{\alpha\mspace{-3mu}-\mspace{-3mu}1}\ln\operatorname{Tr}[H\mspace{-2mu}\rho]\mspace{-1mu}-\mspace{-1mu}\ln\operatorname{Tr}\!\left[ \mspace{-2mu} \left(
H^{\frac{1}{2}}\sigma^{\frac{\alpha-1}{\alpha}}H^{\frac{1}{2}}\right)^{\mspace{-3mu}\frac{\alpha}{\alpha-1}}\mspace{-1mu}\right]\mspace{-1mu}\right\}\!.
\label{eq:sandwiched-variational}
\end{align}
The difficulty in estimating these objective functions has to do with the second terms in \eqref{eq:umegaki-variational-other}--\eqref{eq:sandwiched-variational}, due to noncommutativity. A possible approach for evaluating the second term in \eqref{eq:umegaki-variational} is to employ the Lie--Trotter product formula, which implies 
\begin{equation}
    \operatorname{Tr}
[\exp( H + \ln\sigma)  ] = \lim_{\ell \to \infty} \operatorname{Tr}\!\left[
\left(e^{H/\ell}  \sigma^{1/\ell}\right)^{\ell}  \right].
\label{eq:lie-trotter}
\end{equation}
To realize (a finite approximation of) the right-hand side of \eqref{eq:lie-trotter}, one could employ a variant of multivariate trace estimation~\cite{quek2022multivariate} along with quantum singular value transformation \cite{GSLW19} as a means to realize fractional powers of the density operator $\sigma$ from block encodings of it. This exploration and the associated error analysis is left for future work.

An alternative approach could use the following relations between the measured and  unmeasured quantities \cite[Proposition~4.12]{tomamichel2015quantum}:
\begin{align}
\sD_{M}(\rho\Vert\sigma) & \leq\sD(\rho\Vert\sigma) \leq 
\sD_{M}(\rho\Vert\sigma)+2\ln\left\vert \text{spec}(\sigma)\right\vert ,\\
\sD_{M,\alpha}(\rho\Vert\sigma) & \leq\widetilde{\sD}_{\alpha}(\rho\Vert\sigma)  \leq 
\sD_{M,\alpha}(\rho\Vert\sigma)+2\ln\left\vert \text{spec}(\sigma)\right\vert ,
\end{align}
with the latter holding for all $\alpha\in(0,1)\cup(1,\infty)$. In the above, $\text{spec}(\sigma)$ denotes the set of distinct eigenvalues of $\sigma$. Now, for $n\in \mathbb{N}$, using the fact that $|\text{spec}(\sigma^{\otimes n})| \leq (n+1)^{d-1}$, as well as the additivity of the unmeasured relative entropies, we obtain
\begin{align}
    \left|\frac{\sD_{M}(\rho^{\otimes n}\Vert\sigma^{\otimes n})}{n}  -\sD(\rho\Vert\sigma)\right| \leq 
\frac{2(d-1)}{n}\ln(n+1) ,
\label{eq:err-bnd-1-pinching}\\
\left|\frac{\sD_{M,\alpha}(\rho^{\otimes n}\Vert\sigma^{\otimes n})}{n}  -\widetilde{\sD}_{\alpha}(\rho\Vert\sigma)  \right| \leq 
\frac{2(d-1)}{n}\ln(n+1) .
\label{eq:err-bnd-2-pinching}
\end{align}
One can then rewrite \eqref{eq:rel_ent_variational} and \eqref{eq:renyi_rel_ent_variational} as 
\begin{align}
    &\sD_{M}(\rho^{\otimes n}\Vert\sigma^{\otimes n})  \notag \\
    &\quad =
    \sup_{\omega^{(n)}>0}\left\{\operatorname{Tr}[(\ln \omega ^{(n)})\rho^{\otimes n}]-\operatorname{Tr}[\omega^{(n)}\sigma^{\otimes n}]\right\}+1\!,\label{eq:multi-letter-rel-ent}\\
&\sD_{M,\alpha}(\rho^{\otimes n}\Vert\sigma^{\otimes n}) \nonumber\\
&\quad=\sup_{\omega^{(n)}>0}\left\{\frac{\alpha}{\alpha-1}\ln\operatorname{Tr}[(\omega^{(n)})^{\frac{  \alpha-1}{\alpha} }\rho^{\otimes n}
]-\ln\operatorname{Tr}[\omega^{(n)}\sigma^{\otimes n}]\right\}.
\label{eq:multi-letter-renyi-rel-ent}
\end{align}
It follows from operator concavity of $x\mapsto \ln x$  and $x \mapsto x^{\frac{  \alpha-1}{\alpha}}$ for $\alpha > 1$, the operator convexity of $x \mapsto x^{\frac{  \alpha-1}{\alpha}}$ for $\alpha \in (1/2,1)$, and the permutation invariance of the tensor-power states $\rho^{\otimes n}$ and $\sigma^{\otimes n}$, that in  \eqref{eq:multi-letter-rel-ent}--\eqref{eq:multi-letter-renyi-rel-ent} it suffices to optimize only over permutation invariant observables~$\omega^{(n)}$. This significantly simplifies the optimization since the parameter space occupied by permutation invariant observables grows only polynomially in the number of copies~$n$ (this is related to some recent observations of geometric quantum machine learning \cite{nguyen2022theory,MMGMAWE23}). However, in spite of this reduction, the error bounds in \eqref{eq:err-bnd-1-pinching}--\eqref{eq:err-bnd-2-pinching} scale as $d/n$, and thus the number of copies $n$ must be larger than the dimension $d$ to obtain a good approximation. As $d$ is exponential in the number of qubits, this approach may not be feasible, and new ideas are needed to address this problem.

%\vspace{2mm}

\begin{acknowledgements}

We thank Paul Alsing, Daniel Koch, Saahil Patel, Shannon Ray, and Soorya Rethinasamy for insightful discussions. DP and MMW acknowledge support from the National Science Foundation under Grant No.~1907615.
ZG is partially supported by NSF grants CCF-2046018, DMS-2210368, and CCF-2308446, and the
IBM Academic Award. SS acknowledges support from the Excellence Cluster - Matter and Light for Quantum Computing (ML4Q).

\end{acknowledgements}

\appendix

\section{Connection to Donsker--Varadhan Formula for Relative Entropy and
 Neural Estimation of Relative Entropy}

It is worthwhile to note the connection between the variational formulas for
measured relative entropy in \eqref{eq:rel_ent_variational_other}--\eqref{eq:rel_ent_variational} and the Donsker--Varadhan (DV) formula for
the classical relative entropy of two probability distributions $p$ and $q$ on
$\left\{  1,\ldots,d\right\}  $. For the latter, see \cite[Lemma~2.1]{donsker1975asymptotic} and  \cite[Lemma~1 and Eq.~(8)]{NWJ10}. Based on this connection, we also note here
an additional interpretation of our neural quantum estimation approach for
estimating measured relative entropy. Namely, our approach can be understood
as the parameterized quantum circuit searching for the optimal measurement to
perform and the classical neural network searching for the optimal function in
the DV formula.

To begin with, let us recall the definition of the classical relative entropy:%
\begin{equation}
D(p\Vert q)\coloneqq \sum_{i=1}^{d}p(i)\ln\!\left(  \frac{p(i)}{q(i)}\right)  .
\end{equation}
The DV variational formula for $D(p\Vert q)$ is as follows:%
\begin{align}
D(p\Vert q)  & =\sup_{f}\left\{  \sum_{i=1}^{d}p(i)f(i)-\ln\sum_{i=1}%
^{d}q(i)e^{f(i)}\right\}  \label{eq:DV-var-form-1}\\
& =\sup_{f}\left\{  \sum_{i=1}^{d}p(i)f(i)-\sum_{i=1}^{d}q(i)e^{f(i)}\right\}
+1,\label{eq:DV-var-form-2}%
\end{align}
where the supremum is over all functions $f:\left\{  1,\ldots,d\right\}
\rightarrow\mathbb{R}$. By including an optimization over all rank-one,
projective measurements, the measured relative entropy can thus be written as
follows:%
\begin{equation}
D_{M}(\rho\Vert\sigma)=\sup_{\mathcal{P}}D(p_{\mathcal{P}}\Vert q_{\mathcal{P}%
}),\label{eq:proj-meas-rel-ent}%
\end{equation}
where $\mathcal{P}\coloneqq \left\{  |\phi_{i}\rangle\!\langle\phi_{i}|\right\}
_{i=1}^{d}$ denotes a rank-one, projective measurement (i.e., such that
$\langle\phi_{i}|\phi_{j}\rangle=\delta_{ij}$ and $\sum_{i=1}^{d}|\phi
_{i}\rangle\!\langle\phi_{i}|=I$) and%
\begin{equation}
p_{\mathcal{P}}   \coloneqq \operatorname{Tr}[|\phi_{i}\rangle\!\langle\phi_{i}%
|\rho],\qquad 
q_{\mathcal{P}}   \coloneqq \operatorname{Tr}[|\phi_{i}\rangle\!\langle\phi_{i}%
|\sigma].
\end{equation}
Theorem~2 of \cite{Berta2015OnEntropies} established the fact that it suffices to restrict the optimization in
\eqref{eq:proj-meas-rel-ent} to rank-one, projective measurements. To every
orthonormal basis $\left\{  |\phi_{i}\rangle\right\}  _{i=1}^{d}$, there is a
unitary $U$ that relates it to the computational basis (i.e., $|\phi
_{i}\rangle=U|i\rangle$ for all $i\in\left\{  1,\ldots,d\right\}  $). As such,
the optimization in \eqref{eq:proj-meas-rel-ent} is equivalent to%
\begin{equation}
D_{M}(\rho\Vert\sigma)=\sup_{U}D(p_{U}\Vert q_{U}),
\end{equation}
where%
\begin{equation}
p_{U}   \coloneqq \operatorname{Tr}[|i\rangle\!\langle i|U^{\dag}\rho U],\qquad 
q_{U}   \coloneqq \operatorname{Tr}[|i\rangle\!\langle i|U^{\dag}\sigma U].
\end{equation}
Then plugging into \eqref{eq:DV-var-form-1}--\eqref{eq:DV-var-form-2}, we find
that%
\begin{equation}
\begin{split}
D_{M}(\rho\Vert\sigma)  & =\sup_{U}\sup_{f}\left\{  \sum_{i=1}^{d}%
p_{U}(i)f(i)-\ln\sum_{i=1}^{d}q_{U}(i)e^{f(i)}\right\}
  \\
& =\sup_{U}\sup_{f}\left\{  \sum_{i=1}^{d}p_{U}(i)f(i)-\sum_{i=1}^{d}%
q_{U}(i)e^{f(i)}\right\}  +1.
\end{split}
\label{eq:U-f-meas-rel-ent-2}%
\end{equation}
Let us define%
\begin{equation}
H=\sum_{i=1}^{d}f(i)U|i\rangle\!\langle i|U^{\dag},\label{eq:herm-op-expr}%
\end{equation}
and observe that
\begin{align}
\label{eq:app-proof-1}
\sum_{i=1}^{d}p_{U}(i)f(i)  & =\sum_{i=1}^{d}f(i)\operatorname{Tr}%
[|i\rangle\!\langle i|U^{\dag}\rho U]\\
& =\operatorname{Tr}\left[  \sum_{i=1}^{d}f(i)U|i\rangle\!\langle i|U^{\dag}%
\rho\right]  ,\\
& =\operatorname{Tr}[H\rho]\\
\sum_{i=1}^{d}q_{U}(i)e^{f(i)}  & =\sum_{i=1}^{d}e^{f(i)}\operatorname{Tr}%
[|i\rangle\!\langle i|U^{\dag}\sigma U]\\
& =\operatorname{Tr}\left[  \sum_{i=1}^{d}e^{f(i)}U|i\rangle\!\langle i|U^{\dag
}\sigma\right]  \\
& =\operatorname{Tr}[e^{H}\sigma].
\label{eq:app-proof-last}
\end{align}
Then by inspecting \eqref{eq:U-f-meas-rel-ent-2}, \eqref{eq:herm-op-expr},  and \eqref{eq:app-proof-1}--\eqref{eq:app-proof-last}, and using the fact that every Hermitian operator can
be written as in \eqref{eq:herm-op-expr}, we note that the above is an
alternate derivation of the formulas in \eqref{eq:rel_ent_variational_other}--\eqref{eq:rel_ent_variational}. This alternate
derivation instead uses the DV formula as a starting point. Indeed, one can
also make this observation and connection by inspecting the proof of \cite[Lemma~1]{Berta2015OnEntropies}. 

Now comparing \eqref{eq:U-f-meas-rel-ent-2} and \eqref{eq:meas_rel_ent_parameterized}--\eqref{eq:estterms2}, we see that the
parameterized optimization in \eqref{eq:meas_rel_ent_parameterized} is equivalent to optimizing over parameterized unitaries
in order to find the optimal measurement and optimizing over parameterized functions
according to the DV formula. Thus, our development here shows that our
approach for quantum neural estimation of the measured relative entropy is
equivalent to using classical neural estimation according to the DV formula
combined with a parameterized quantum circuit to find the optimal measurement.

\bibliography{references}

\end{document}